\documentclass{aa} 

\usepackage[breaklinks=true]{hyperref} 

\usepackage[varg]{txfonts}
\usepackage{multirow}
\usepackage{color}
\usepackage{ulem}

\usepackage{graphicx}
\usepackage{subcaption}
\graphicspath{{./DMC/}}

\usepackage{natbib}
\bibpunct{(}{)}{;}{a}{}{,} 

\usepackage{comment}

\renewcommand{\d}{\mathrm{d}}
\renewcommand{\L}{L}

\newcommand{\UnitMd}{\,\mathrm{M_\odot\, yr^{-1}}}
\newcommand{\UnitV}{\,\mathrm{km\,s^{-1}}}

\newcommand{\UnitT}{\,\mathrm{kK}}
\newcommand{\Unitkap}{\,\mathrm{cm^2\,g^{-1}}}
\newcommand{\Unittime}{\,\mathrm{ks}}

\newcommand{\opal}{{\rm OPAL }}
\newcommand{\OPAL}{{\rm OPAL}}
\newcommand{\cak}{{\rm CAK }}
\newcommand{\CAK}{{\rm CAK}}
\newcommand{\tot}{{\rm tot}}
\newcommand{\esc}{{\rm esc}}

\newcommand{\p}{\partial}

\definecolor{coathA_color}{RGB}{22, 82, 145}
\definecolor{coathB_color}{RGB}{16, 110, 23} 
\definecolor{coathC_color}{RGB}{141, 143, 53}
\definecolor{coathD_color}{RGB}{122, 25, 20}
\definecolor{newtxt_color}{RGB}{0, 0, 255}
\definecolor{oldtxt_color}{RGB}{255, 0, 0}

\newcommand{\newtxt}[1]{{#1}}
\newcommand{\oldtxt}[1]{\iffalse {#1} \fi}

\begin{document}

\title{\newtxt{Dynamically inflated wind models of classical Wolf-Rayet stars}
\oldtxt{Wind models of dynamically inflated classical Wolf-Rayet stars}}
\author{L. G. Poniatowski\inst{1} 
            \and 
        J. O. Sundqvist\inst{1}
            \and
        N. D. Kee\inst{1}
            \and
        S. P. Owocki\inst{2}
            \and
        P. Marchant\inst{1}
            \and
        L. Decin\inst{1}
            \and
        A. de Koter\inst{1,3}
            \and
        L. Mahy\inst{1}
           \and
        H. Sana\inst{1}
         }
\institute{Institute of Astronomy, KU Leuven, 
           Celestijnenlaan 200D, 3001, Leuven, Belgium
              \and
           Bartol Research Institute, Department of Physics and Astronomy,
           University of Delaware, Newark, DE 19716, USA 
              \and
           Astronomical Institute Anton Pannekoek, Amsterdam University, 
           Science Park 904, 1098 XH Amsterdam, The Netherlands}
              
\date{\today}

 \abstract
{Vigorous mass loss in the classical Wolf-Rayet (WR) phase is important for the late evolution and final fate of massive stars.} {We develop spherically symmetric time-dependent and steady-state \oldtxt{radiation-}hydrodynamical models of the \newtxt{radiation-driven} wind outflows and associated mass loss from classical WR stars.}
{The simulations are based on combining the opacities typically used in static stellar structure and evolution models with a simple parametrised form for the enhanced line-opacity expected within a supersonic outflow.} {Our simulations reveal high mass-loss rates initiated in deep and hot optically thick layers around $T \approx 200\,\rm kK$. The resulting velocity structure is non-monotonic and can be separated into three phases: {\it i)} an initial acceleration to supersonic speeds (caused by the static opacity), {\it ii)} stagnation and even deceleration, and {\it iii)} an outer region of rapid re-acceleration (by line-opacity). The characteristic structures seen in converged steady-state simulations agree well with the outflow properties of our time-dependent models.} 
{By directly comparing our dynamic simulations to corresponding hydrostatic models, we demonstrate explicitly that the need to invoke extra energy transport in convectively inefficient regions of stellar structure and evolution models (in order to prevent drastic inflation of static WR envelopes) is merely an artefact of enforcing a hydrostatic outer boundary. Moreover, the "dynamically inflated" inner regions of our simulations provide a natural explanation for the often-found mismatch between predicted hydrostatic WR radii and those inferred from spectroscopy; by extrapolating a monotonic $\beta$-type velocity law from the "observable" supersonic regions to the invisible hydrostatic core, spectroscopic models likely overestimate the core radius by a factor of a few. Finally, we contrast our simulations with alternative recent WR wind models based on co-moving frame (CMF) radiative transfer for computing the radiation force. Since CMF transfer currently cannot handle non-monotonic velocity fields, the characteristic deceleration regions found here are avoided in such simulations by invoking an ad-hoc very high degree of clumping. 
}


\titlerunning{Dynamically inflated wind models of classical Wolf-Rayet stars}

\maketitle

\section{Introduction}\label{Se_introduction}

The evolution of stars with initial masses higher than eight times that of the Sun plays an essential role in the chemistry and dynamics of galaxies like our Milky Way \citep{Crowther_07, Doran_13, Ramachandran_18, Prantzos_18}. These massive stars are a vital source of heavy elements and UV radiation, enriching their surroundings through strong radiation-driven winds  \citep{Lucy_70, Castor_75, Puls_08, Vink_01, Bjorklund_20}. In the final evolutionary stages, some of those massive stars become hydrogen-depleted, typically core He-burning progenitors of neutron stars or black holes \citep{Yoon_12, Groh_13b}. Among these are so called classical Wolf-Rayet (WR) stars \citep{Wolf_1867}, which are characterised by strong spectral emission lines and a high luminosity to mass ratio, $\L/M_\star\sim10^{4}\L_\odot/M_\odot$, \citep{Crowther_07}. Classical WR stars are distinct from very massive, main-sequence WR stars, which are core H-burning \citep{deKoter_97,Crowther_10}. Classical WR stars are also different in their composition and evolutionary state from hydrogen-deficient WR central stars of planetary nebula \citep[e.g.][]{Todt_10}.

As first suggested by \citet{Beals_29}, the prominent emission lines visible in WR spectra indicate strong stellar winds with high terminal speeds ($v_\infty\sim 2000\dots 3000\UnitV$) and mass-loss rates  ($\dot M\sim 10^{-5}\dots10^{-3}~\UnitMd$) \citep{Hamann_19, Sander_19}. However, while the overall wind properties of massive main-sequence OB-stars (see \citealt{Puls_08} for a review) are 
quite well reproduced by the  line-driven wind theory formulated first by \citet*{Castor_75} (\CAK), this standard theory typically fails to explain the order of magnitude higher mass-loss rates of classical WR stars \citep{Cassinelli_91,Lamers_93}. 

Nonetheless, WR winds are still thought to be radiation-driven, as their high $\L/M_\star$ ratio brings them \oldtxt{generally} close to the limit at which the acceleration due to radiation $g_{\rm r}$ balances that of gravity $g$ (i.e. close to $\Gamma = 1$, for "Eddington factor" $\Gamma = g_{\rm r}/g$). 
\newtxt{In fact, consulting the opacities used in stellar structure calculations which are}
\oldtxt{Opacities in stellar structure computations are} generally obtained from tabulations computed assuming a static medium \citep[e.g.,][]{Iglesias_96}, \oldtxt{and using such tables} the Eddington limit for classical WR stars is \newtxt{already reached} \oldtxt{reached already} in sub-surface layers with temperatures of about $150-250\UnitT$. 
Since convection is highly inefficient in these layers \citep{Grafener_12}, static models then typically display density inversions and highly inflated stellar envelopes \citep{Ishii_99, Petrovic_06, Grafener_12, Sanyal_15}. For corresponding stellar evolution calculations, such envelope-inflation \newtxt{is computationally difficult to treat} \oldtxt{presents great computational challenges} and various numerical "tricks" are thus typically required to make computations tractable \citep{Paxton_13,Ekstrom_12}. This then leads to very hot ($T_{\rm eff} \ga 100\UnitT$) and compact ($R \sim R_\odot$) WR stellar surfaces. On the other hand, spectroscopic studies aiming to constrain the WR surface from \newtxt{observationally} inferred $T_{\rm eff}$ typically deduce hydrostatic radii that are factors of $\sim 2-3$ higher than predicted by such evolution models \citep[e.g.,][]{Crowther_07,Sander_19,Hamann_19}; this mismatch is sometimes referred to as the "core-radius" problem of classical WR stars.

\newtxt{Rather} \oldtxt{However, rather} than retaining a static envelope, breaching the Eddington limit in sub-photospheric layers can initiate an optically thick supersonic outflow. 
\citet{Nugis_02} and \citet{Grassitelli_18} suggested that WR mass-loss rates can be computed simply by considering only the conditions at the point where the sonic speed is reached (at $\Gamma \approx 1$, since gas pressure terms typically are very small in comparison). Using the \newtxt{\opal tables of Rosseland mean opacity} \oldtxt{\opal opacity tables} \citep{Iglesias_96}, \citet{Ro_16} showed that supersonic velocities are indeed found in deep sub-photospheric layers, but that a successful wind solution that could bring the initiated mass flux to infinity could not be found. 

\newtxt{However, these models neglect} \oldtxt{These models neglect, however,} the strong enhancement of the line-opacity expected in a supersonic outflow. Previous modeling attempts including the Doppler effect in the line opacity calculations have either relied on a pre-assumed fixed velocity field to solve for the mass loss \citep[e.g., see][]{Lucy_93,Springmann_94,Springmann_98,deKoter_97}, or attempted an iterative (assuming time-independence) solution toward a self-consistent velocity field and mass loss \citep[][]{Grafener_05,Sander_20a,Sander_20b}. The latter models have used comoving-frame (CMF) radiative transfer for the calculation of $g_r$. Such CMF transfer is a computationally intensive numerical technique that requires the velocity field to remain smooth and monotonic (see, e.g., discussion in \citealt{Sander_20a}). Moreover, all these previous studies of classical WR outflows have been performed in the steady-state limit. 

In this paper we present a first attempt to build a model that addresses both  (dynamic) envelope-inflation and  line-driving, using a "hybrid" opacity approach based on combining the opacities used for static stellar structure calculations with a simple variant of the standard parameterisation for line-opacities in supersonic flows. \newtxt{This formalism then allows for computation of both steady-state and time-dependent WR wind structures.} The organization of the paper is as follows: 
In Section \ref{Se_phymodel}, we describe our basic physical \newtxt{set-up} \oldtxt{setup}. We present dynamical models in the steady-state limit and compare these to corresponding static calculations in Section \ref{Se_steady_state}. Then in Section \ref{Se_Time_dependant} we compare these steady-state models to full time-dependent radiation-hydrodynamical simulations of "dynamically inflated" WR outflows. In Section \ref{Se_discussion} we discuss our results and some open questions regarding WR stellar outflows, and finally we summarise our results and provide an outlook for future work in Section \ref{Se_summary_future}.

\section{Physical model}\label{Se_phymodel}

We describe the WR wind outflow by the appropriate hydrodynamical equations of mass and momentum conservation assuming spherical symmetry:
\begin{equation}\label{Eq_time_continuity}
\frac{\p\rho}{\p t} + \frac{1}{r^2}\frac{\p }{\p r}(r^2\rho v) = 0\,,
\end{equation}
\begin{equation}\label{Eq_time_momentum_full}
\frac{\p}{\p t}(\rho v) + \frac{1}{r^2}\frac{\p}{\p r}(r^2v\rho v ) = -\frac{\p P_{\rm g}}{\p r}  -\rho g + \rho g_{\rm r}\,. 
\end{equation} 
Here $\rho$, $v$ and  $P_{\rm g} = k_{\rm B} T \rho / (\mu m_{\rm H})$ are mass density, velocity and gas pressure, where $T$ is gas temperature, $k_{\rm B}$ is the Boltzmann constant, $\mu$ \newtxt{is the} mean molecular weight, and $m_{\rm H}$ is the mass of the hydrogen atom, $g =  G M_\star/r^2$ is the gravitation acceleration for a constant stellar mass $M_\star$ and 
\begin{equation}\label{Eq_grad_general}
g_{\rm r} = \frac{\kappa_{\rm F} {F}} {c} \,,
\end{equation}
is the acceleration due to stellar radiation, for radiation flux ${F}$ and flux weighted opacity  ("mass absorption coefficient") $\kappa_{\rm F}$ in $\Unitkap$.
Using the Eddington factor 
we can also write the equation of motion (e.o.m) as: 
\begin{equation}\label{Eq_time_momentum}
\frac{\p}{\p t}(\rho v) + \frac{1}{r^2}\frac{\p}{\p r}(r^2v\rho v ) = -\frac{\p P_{\rm g}}{\p r}  -\rho\frac{M_\star G}{r^2}(1 - \Gamma)\,. 
\end{equation} 
%
The Eddington ratio can be expressed as:
\begin{equation}
\Gamma = \frac{\kappa_F\L}{4\pi M_\star G c}\,,
\end{equation} 
for a stellar luminosity $\L = 4 \pi r^2 F$. In this paper, we assume that the radiative luminosity remains constant throughout the outflow. This means \newtxt{that,} we neglect a term corresponding to the work of the radiation field against gravity. We can estimate the corresponding expected luminosity variation 
by computing the photon-tiring parameter $m = {\dot M}/{\dot M_{\max}}$ \citep{Owocki_17}, which is the ratio of the stellar mass loss $\dot M$ to the maximum amount of mass loss that the stellar luminosity can drive, $\dot M_{\max} = \L R_{\rm c}/(M_\star G)$, where $R_{\rm c}$ is the "core radius" defined at the fixed lower boundary. Taking the stellar parameters used throughout this paper, $M_\star = 10M_\odot,$ $R_{\rm c}= 1R_\odot$, $\log_{10}(\L/\L_\odot) = 5.416$, and a typical order of magnitude mass loss, $\dot M \sim 5.0\,\cdot 10^{-5}\UnitMd$, 
one gets
\begin{equation}
m =\frac{\dot M GM_\star}{R_{\rm c}\L}\approx 0.06\,,
\end{equation} 
which demonstrates that the luminosity variation is only a marginal effect for the cases considered in this paper.

\newtxt{In practice, especially for time-dependent dynamical computations (with possibly non-monotonic velocity fields) it is not computationally feasible to derive the temperature structure, e.g., from radiative equilibrium by means of full solutions to the frequency dependent radiative transfer equations. To simplify,
we therefore follow the common approach of replacing the full energy equation by the \citet{Lucy_71} analytic radiative equilibrium model for a grey, spherically symmetric, diluted atmosphere, however, replacing the grey opacity by the actual flux weighted $\kappa_{\rm F}$ \citep[e.g., see][]{Lucy_93}. This allows us to write the temperature structure as:}
\oldtxt{To compute the temperature structure, we replace the full energy equation by the citet{Lucy\_71} radiation transport model for a spherically symmetric diluted atmosphere }
\begin{equation}\label{Eq_Lucy}
T^4 = T_{\rm c,\,eff}^4\left(W(r) + \frac{3}{4}\tau_{\rm sp}\right)\;,
\end{equation}
%
where $T_{\rm c,\,eff}^4 = \L/(4\pi\sigma_{\rm SB}R_{\rm c}^2)$ is a "core effective temperature", defined here from the lower boundary set at $R_c$, $\sigma_{\rm SB}$ \newtxt{is} the Stefan-Boltzmann constant, $W(r) = 0.5\left(1-\sqrt{1-R_{\rm c}^2/r^2}\right)$ \newtxt{is} the dilution factor, and $\tau_{\rm sp}$ \newtxt{is} the spherically modified optical depth
\begin{equation}\label{Eq_tau_sp}
\d\tau_{\rm sp} = -\kappa_{\rm F} \rho\left(\frac{R_{\rm c}}{r}\right)^2\d r\;,
\quad\text{or}\quad
\tau_{\rm sp} = \oldtxt{-}\int_{r}^{\infty}\d r^\prime\kappa_{\rm F} \rho\left(\frac{R_{\rm c}}{r^\prime}\right)^2\;.
\end{equation}
Here again the radius of our fixed lower boundary $R_{\rm c}$ is used as a scaling radius for the spherically modified optical depth scale. This is similar to the WR models by \citet{Sander_20a} who also define their stellar radius from the hydrostatic lower boundary. It is different, however, than, e.g., \citet{Sundqvist_19} who set the reference radius for the temperature structure in their O-star models at a photospheric radius $R_{\rm ph}$.
An argument against using $R_{\rm ph}$ as a scaling radius for optically thick WR winds is, however, that $R_{\rm ph} \gg R_{\rm c}$, such that the geometric dilution of the radiation field would be neglected for large parts of the outflow \oldtxt{if $R_{\rm ph}$ was used as the scaling radius} \citep[see also discussion in ][]{Nugis_02}. In principle, the explicit choice of the reference radius will affect the temperature structure throughout our models. However, test-calculations have shown that this is a quite marginal effect for the resulting mass loss and velocities of the simulations considered in this paper. In any case, to facilitate comparison with other models we also introduce a "stellar photospheric effective temperature" $T_{\rm  ph,\,eff}$ at this photospheric radius $R_{\rm ph}$. \newtxt{Since} \oldtxt{However, since} the  definition of a stellar photosphere in a spherically diluted stellar envelope is non-trivial\footnote{For example, in Section \ref{Se_steady_state_static_inflation} the stellar photoshere is located at the $\tau_{\rm sp} = 4/3$ surface in the radiation diffusion approximation, rather than the typically assumed photoshere location at the $\tau_{\rm sp} = 2/3$ surface.} and depends on the specific computation of the temperature structure, we here simply approximate
\begin{equation}\label{Eq_Tpheff}
    T(R_{\rm ph})^4  = \frac{\L}{4\pi\sigma_{\rm SB}R_{\rm ph}^2} \equiv T_{\rm ph,\, eff}^4\,,
\end{equation}
such that the photospheric and core effective temperatures and radii are related through
\begin{equation}\label{Eq_Tceff_to_Teff}
    T_{\rm ph,\,eff} = \sqrt{\frac{R_{\rm c}}{R_{\rm ph}}}T_{\rm c,\, eff}\,.
\end{equation}

For given stellar parameters $\L$, $M_\star$, and $R_{\rm c}$, and a known variation of $\kappa_{\rm F}$, Eqs. \ref{Eq_time_continuity}-\ref{Eq_tau_sp} form the basic system of coupled differential equations under investigation in this paper. In general, $\kappa_{\rm F}$ is a complicated function that depends on chemical composition, density, temperature, velocity and redial position of the absorbing and emitting gas.
In this study, we focus on two cases. First, we consider static opacities such as those used for stellar structure and evolution computations \citep[see][]{Ro_16}. Due to the high continuum optical depth of WR outflows, conditions in the wind launching regions resemble those of radiation diffusion, allowing us to replace the flux-weighted opacity with a Rosseland mean, and so to utilise standard \opal tables for the opacity-mapping as a function of $T$ and $\rho$ \citep{Iglesias_96}. However, since these tabulations ignore the influence of Doppler shifts on the line-opacity  \citep[which are critical for WR wind driving, e.g.,][]{Sander_20a} we will also consider a "hybrid-model" accounting at least approximately for this effect. More specifically, we will use a \CAK-like parametrisation to add the cumulative force from an ensemble of spectral lines to the \opal tables (see, \newtxt{Section} \oldtxt{Sect.} \ref{Se_steady_state_faild_wind}).

\section{Steady-state approximation}\label{Se_steady_state}
To build physical insight we first examine a simplified steady-state case.
\newtxt{Equations}\oldtxt{Eqs.} \ref{Eq_time_continuity} and \ref{Eq_time_momentum} are \newtxt{then} written as:
\begin{equation}\label{Eq_massloos_rate}
\dot M  = 4\pi r^2 \rho v = \text{Constant}\,,
\end{equation}
\begin{equation}\label{Eq_motion}
\left(1-\frac{a^2}{v^2}\right)v\frac{\d v}{\d r} = \frac{2a^2}{r} - \frac{\d a^2}{\d r} - \frac{M_\star G}{r^2}(1 -\Gamma)\,,
\end{equation}
for an isothermal sound speed $a^2 = k_{\rm B} T/(\mu m_{\rm H})$.
Equation \ref{Eq_massloos_rate} defines the spherically symmetric mass loss rate $\dot M$. Here we assume $\mu = 4/3$ corresponding to a fully ionised helium plasma.

It is commonly believed that the winds of  WR stars are initiated in the deep sub-photospheric layers where the gas temperature is $T \sim 150-200\UnitT$. In this temperature region, due to iron recombination, a significant increase of the Rosseland mean opacity is observed, sometimes called the "iron opacity bump". A simple comparison of scales reveals that \newtxt{throughout the supersonic wind outflow} the first two terms of the r.h.s. in Eq. \ref{Eq_motion} are much smaller than the radiation and gravitation terms, so that we can approximate the e.o.m with
\begin{equation}\label{Eq_motion_simp}
\frac{1}{2}\left(1 - \frac{a^2}{v^2} \right)\frac{\d v^2}{\d r} = \frac{M_\star G}{r^2}(\Gamma - 1)\;. 
\end{equation}
Here, the sound-speed term on the l.h.s. is retained in order to enable a mapping of the supersonic solution onto the subsonic part. 

\subsection{Boundary conditions}\label{Se_steady_state_twopointer_boundary}

Equations \ref{Eq_Lucy}-\ref{Eq_motion_simp}
impose a two-point boundary value problem, with boundaries at $R_{\rm c}$ and  at\footnote{for numerical computations the outer formal boundary is simply replaced by 
a maximum radius $R_{\max}\gg R_{\rm c}$} $r\rightarrow\infty$. 

\textit{The inner boundary} at $r = R_{\rm c}$ is defined to be a sonic-point, with $v^2 = a^2 = a_{\rm s}^2$, \newtxt{where $a_{\rm s}$ is defined to be the sound-speed at this sonic-point}.
As can be seen from Eq. \ref{Eq_motion_simp},
this implies $\Gamma(R_{\rm c}) = 1$. Below the sonic point \newtxt{as the velocity rapidly diminishes from its sonic point value $a_{\rm s}$ inwards,} the density structure is well approximated by the hydrostatic solution \citep[e.g. see][]{Ro_16,Grassitelli_18}. In contrast, in the super-sonic part, the density structure diverges from the structure given by the hydrostatic solution. As such, we initiate the calculation at the sonic-point and only consider the super-sonic part of the outflow, assuming that the sub-sonic region still can be described by hydrostatic equilibrium.
The tabulated \opal opacities can then be used to locate points for which $\Gamma = 1$, constraining the lower boundary density as a function of temperature. In this way, the range of possible mass-loss rates $\dot M  = 4\pi R_{\rm c}^2\rho(a_{\rm s})a_{\rm s}$ 
is set as a function of the lower boundary temperature. 

\textit{The outer boundary} is set by requiring that the wind temperature decreases sufficiently at large radii. Replacing density with the mass-loss rate using Eq. \ref{Eq_massloos_rate}, the spherically modified optical depth (Eq. \ref{Eq_tau_sp}) is used to compute the temperature structure according to Eq. \ref{Eq_Lucy}. We compute the optical depth at the outer boundary $\tau_{\rm out}$ by radially integrating Eq. \ref{Eq_tau_sp} in $r\in[R_{\max},~\infty)$, assuming a terminal velocity $v(r \geq R_{\max}) = v(R_{\max}) = v_\infty$ and $\kappa_{\rm F} (r \geq R_{\max}) = \kappa_e$, where $\kappa_e$ is the Thompson scattering mass absorption coefficient. 
The outer boundary temperature is then set by the maximum between applying $\tau_{\rm out}$ in Eq. \ref{Eq_tau_sp} and a floor value, which is typically set to a few tenths of the stellar effective temperature \citep[see][]{Puls_05,Sundqvist_19}. By matching the inner and outer boundary constraints, we uniquely constrain the wind structure. In turn this then sets the unique mass-loss rate of the model from the range of possible solutions at the lower boundary.

\subsection{Energy requirement}\label{Se_steady_state_energy}

Additionally, for wind material to escape the stellar potential, the global energy requirement must be satisfied, which means that the integrated mechanical energy of the wind has to be positive. Taking the supersonic limit $a^2/v^2\ll 1$ in Eq. \ref{Eq_motion_simp} yields: 
\begin{equation}
    \frac{v_{\infty}^2}{2} -\frac{v_0^2}{2}= \frac{v_\esc^2}{2}\int_{R_{\rm c}}^{\infty}\d r\frac{R_{\rm c}\Gamma}{r^2}  - \frac{v_\esc^2}{2} \,.
\end{equation}
Here $v_0\sim a_{\rm s}$ is the initial velocity and \newtxt{$v_\esc^2 = 2M_\star G/R_{\rm c}$} \oldtxt{$v_\esc^2 = 2MG/R_{\rm c}$} is the escape velocity from $R_{\rm c}$. The first term on the r.h.s. corresponds to the work done by the stellar radiation and the second term is the work done by gravity. We can then introduce $\mathcal{W}$ as the net energy change in units of $v_\esc^2/2$:
\begin{equation}\label{Eq_net_en}
    \mathcal{W}= \int_{R_{\rm c}}^{\infty}\d r\left(\frac{R_{\rm c}\Gamma}{r^2}\right) - 1\,.
\end{equation}
For the wind to escape to $r \rightarrow \infty$, this net energy change has to be positive, i.e. $\mathcal{W}\geq 0$. Alternatively, we can also introduce the \newtxt{ratio of kinetic to core potential energy as a} square of the velocity to core escape velocity ratio:
\begin{equation}\label{Eq_kin_en}
    w(r) = \left(\frac{v(r)}{v_\esc}\right)^2\,,
\end{equation}
which goes to $\mathcal{W}$ as $r\rightarrow\infty$ so that by construction we require $w(r \rightarrow \infty) > \newtxt{0}\oldtxt{1}$ in order to escape. 

\subsection{Failed Winds in the static opacity limit}\label{Se_steady_state_faild_wind}

\begin{figure}[ht!]
\centering
    \begin{subfigure}[t]{.48\textwidth}
    \centering
        \includegraphics[width=\linewidth]{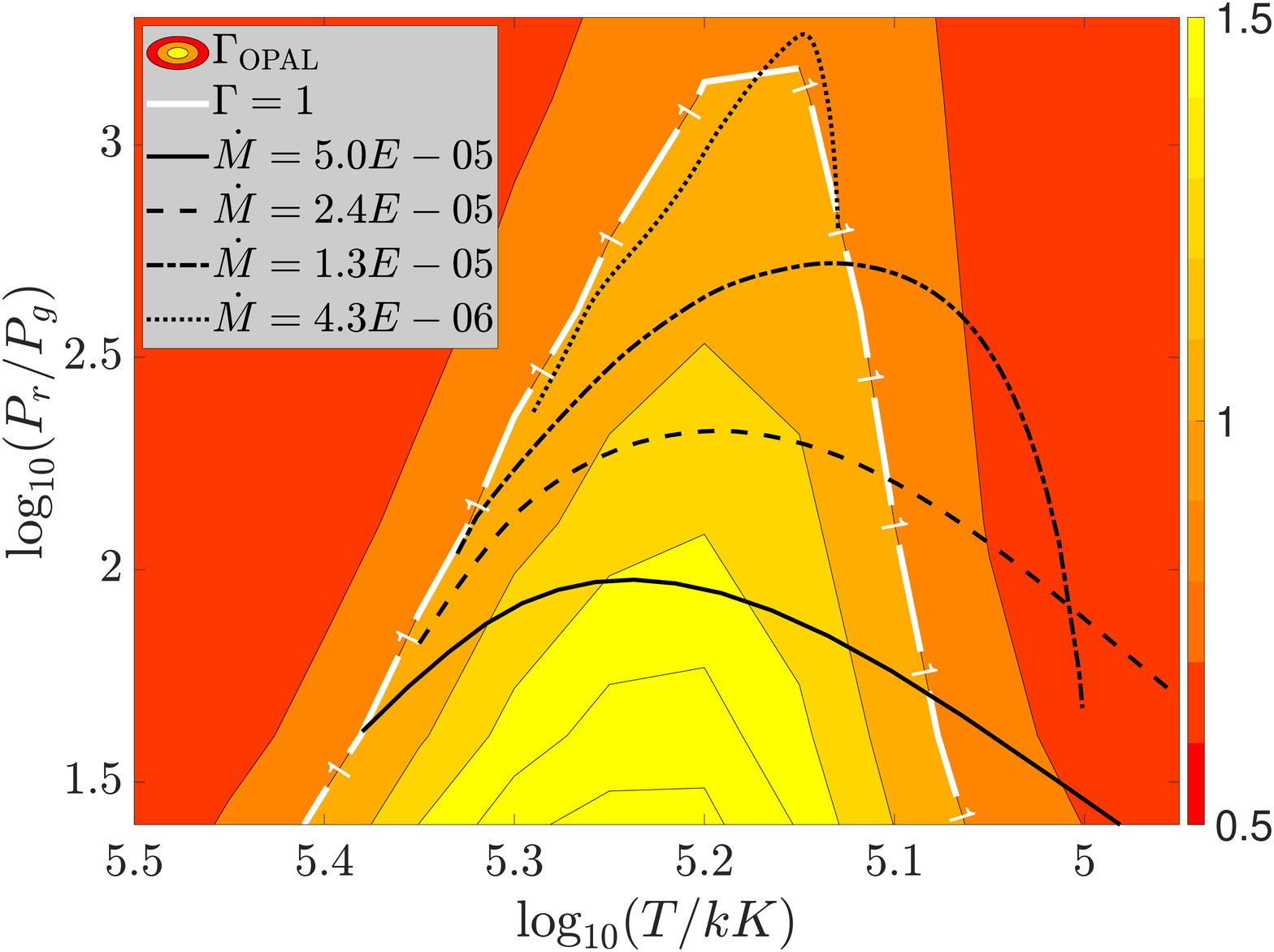}
        \caption{}
        \label{Fig_Ro_comp_opacity}
    \end{subfigure}\\%
    \begin{subfigure}[t]{.48\textwidth}
    \centering
        \includegraphics[width=\linewidth]{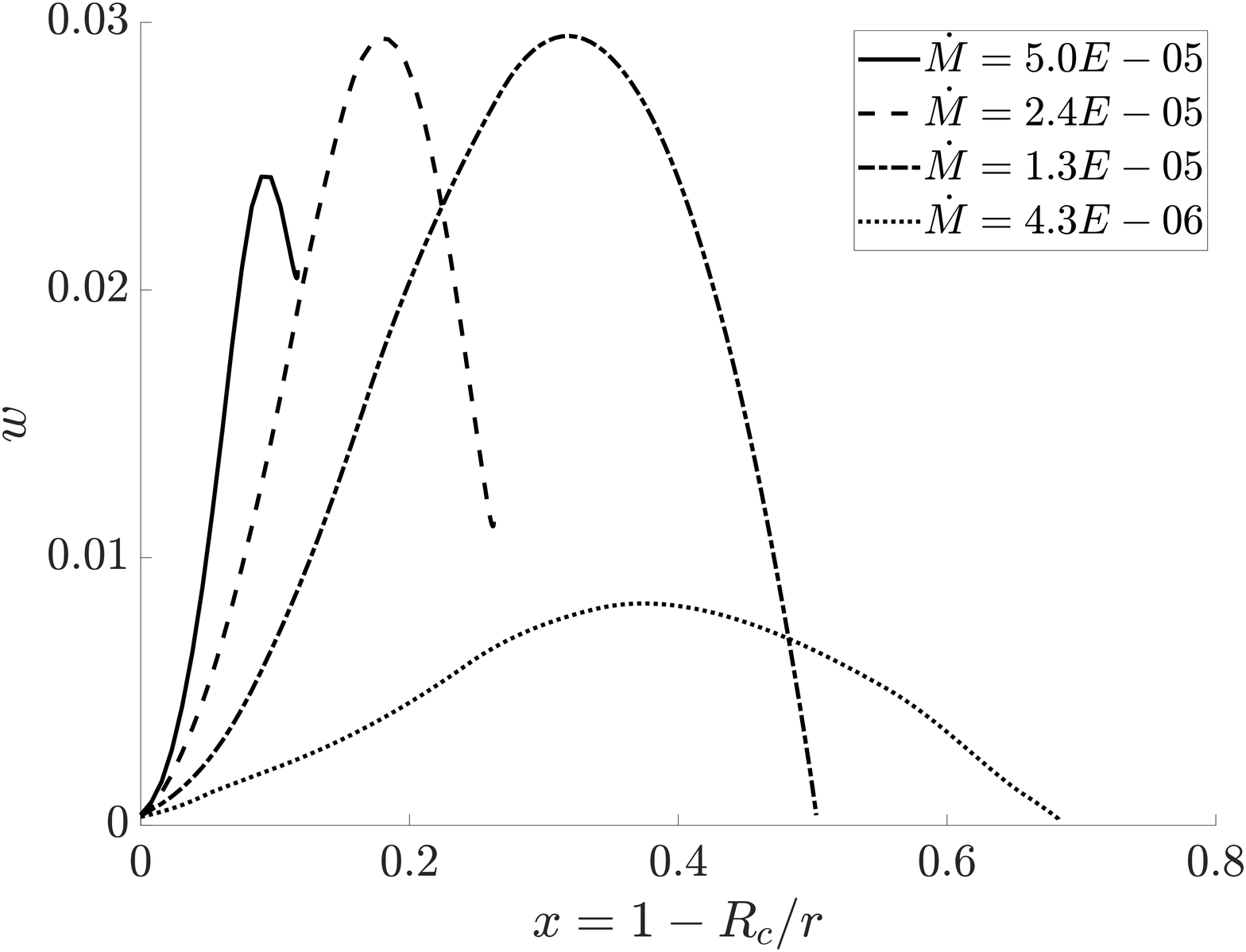}
        \caption{}
        \label{Fig_Ro_comp_energy}
    \end{subfigure}
    \caption{\textit{(a)} Colour map of $\Gamma_\opal$ for the stellar parameters from Table \ref{Tab_stellar_params}. Temperature is given on abscissa and radiation-to-gas pressure ratio on ordinate. $\Gamma(\rho_{\rm s},a_{\rm s}^2) = 1$ is identified with  white contour. \newtxt{The colour bar on the right corresponds to the colour-coding of the plot and gives the numerical value of $\Gamma_\opal$ for a given temperature and radiation-to-gas pressure.} Different line-styles show the radiation-to-gas pressure ratio and temperature structure for the different fixed mass-loss rates.
    \textit{(b)} Ratio of kinetic to core potential energy $w = v^2/v_\esc^2$ of the corresponding solutions from panel (a). The abscissa here shows the radius coordinate $x$. }
\end{figure}

\newtxt{Let us first investigate the possibility of driving the stellar outflow in the static opacity limit, i.e. the limit in which we only consider \opal opacities, thus neglecting the Doppler effect; $\kappa_{\rm F} \approx \kappa_\opal$  \citep[see also][]{Ro_16}. To set up the problem, we choose the stellar mass and the core radius (as given in Table \ref{Tab_stellar_params}) such that they roughly represent the average value of the mass-to-radius ratio of the observationally inferred WR stellar population from \citet{Hamann_19} and \citet{Sander_19}. The luminosity is then set such that the resulting Thompson scattering Eddington ratio $\Gamma_e=0.40$, which also corresponds approximately to the observationally inferred average \citep{Hamann_19,Sander_19}. All models presented in Sections \ref{Se_steady_state} and \ref{Se_Time_dependant} of this paper use these stellar parameters, and also the \opal tables for solar metallicity $Z=0.02$ and composition by \citet{Grevesse_93}. This set-up then gives a range of possible sonic-point values of $a_{\rm s}^2,\:\rho_{\rm s}$ such that the condition $\Gamma(a_{\rm s}^2,\rho_{\rm s}) = 1$ is fulfilled (see Fig. \ref{Fig_Ro_comp_opacity}). These points are used as a single point boundary condition, after which forward integration from the sonic-point is carried out. The results of these integrations for various fixed mass-loss rates $\dot{M} = 4 \pi R_{\rm c} \rho_{\rm s} a_{\rm s}$ are shown in Fig. \ref{Fig_Ro_comp_opacity}.}
\oldtxt{Let us first follow an approach by only considering \opal opacities, thus neglecting the Doppler effect; $\kappa_{\rm F} \approx \kappa_\opal$  citep[see also][]{Ro\_16}. For given stellar parameters, the \opal tables with solar composition composed by citet{Grevesse\_93} and metallicity $Z=0.02$ then give a range of possible sonic-point values of $a_{\rm s}^2,\:\rho_{\rm s}$ such that $\Gamma(a_{\rm s}^2,\rho_{\rm s}) = 1$ (see Fig. \ref{Fig_Ro_comp_opacity}). These points are used as a single point boundary condition, after which forward integration from the sonic-point is carried out. Using our standard stellar parameters from Table \ref{Tab_stellar_params} the results of these integrations for various fixed mass-loss rates $\dot{M} = 4 \pi R_{\rm c} \rho_{\rm s} a_{\rm s}$ are shown in Fig. \ref{Fig_Ro_comp_opacity}.} The figure displays a colour map of $\Gamma$ with the temperature on the abscissa and the ratio between the radiation pressure 
\begin{equation}\label{Eq_rad_pressure}
    P_r = \frac{4 \sigma_{\rm SB}}{3c} T^4\,,
\end{equation}
and gas pressure $P_{\rm g}$ on the ordinate (the ratio $P_{\rm r}/P_{\rm g}$ is a proxy for the density). The integration-curves show the density and temperature structure for the different cases, starting from various positions on the $\Gamma = 1$ curve at the inner boundary $R_{\rm c}$ and extending outwards. Figure \ref{Fig_Ro_comp_energy} then shows the corresponding net gain of kinetic energy (Eq. \ref{Eq_kin_en}), demonstrating that after an initial acceleration and increasing velocities, all of the curves start to decelerate until eventually reaching zero velocity and so terminating the outward integration (some solutions had to be formally terminated before actually reaching zero velocity, due to numerical difficulties in the steep deceleration region). This indicates that none of the potential solutions starting from the $\Gamma =1$ curve \newtxt{are} \oldtxt{is} able to escape the stellar gravitation potential. These results are consistent with \citet{Ro_16}, who also found that \opal opacities were not able to sustain a radiation-driven mass loss initiated in the deep layers around the iron-bump, and indicates that an additional source of opacity is required, raising the question of a  "missing force".

\begin{table}[ht]
    \centering
    \caption{Summary of stellar parameters used in this paper. Here we set our standard stellar parameters such that they roughly represent the \oldtxt{observationally inferred} average value of the mass-to-radius ratio of the \newtxt{observationally inferred} WR stellar population \newtxt{from} \oldtxt{by} \citet{Hamann_19,Sander_19}. The luminosity is then set such that the resulting $\Gamma_e = 0.40$, also corresponding to the observationally inferred average \citep{Hamann_19,Sander_19}.}
    \begin{tabular}{ c c c c c }
         \hline
         $M_\star/M_\odot$ & $R_{\rm c}/R_\odot$ & $\log (\L/\L_\odot) $ & $v_\esc\,(\UnitV)$ \\
         \hline
         $10$ & $1$ & $5.416$ &  $1950$ \\
         \hline
    \end{tabular}

    \label{Tab_stellar_params}
\end{table}

\subsection{Missing force}\label{Se_steady_state_missing_force}

The natural candidate for the missing opacity source discussed above comes from the cumulative effect of spectral lines due to the Doppler effect. To estimate this we use the \cak parametrisation for a distribution of spectral lines\newtxt{, using here a simple approximation of taking the absolute value of the velocity gradient}:
 \begin{equation}\label{Eq_cak_g}
 \Gamma_\cak  = \Gamma_e \frac{\bar Q ^{1-\alpha}}{1-\alpha}\left(\frac{1}{\kappa_e c \rho}\left|\frac{\d v}{\d r}\right|\right)^\alpha\,,
 \end{equation}
where \oldtxt{$\Gamma_e$ is the Thompson scattering Eddington ration,} $\bar Q$ is an effective strength of the line ensemble \citep{Gayley_95}, and $\alpha$ sets the \cak  line distribution power index. \newtxt{In reality, once regions where ${\d v}/{\d r}<0$ arise, it is possible for a photon proceeding from the stellar core to be Doppler shifted into resonance with the same spectral line at multiple locations in the outflow, thereby requiring a fully non-local computation of the line acceleration. As discussed further below, this could (in principle) be handled in a more self-consistent manner by e.g. Monte-Carlo (MC) line-force calculations. 
In this initial study, however, we opt for a simpler approach, in order to make time-dependent simulations feasible as well. Two possible approximations that can be taken then are to either take the absolute value of the velocity gradient $\left|{\d v}/{\d r}\right|$ or use $\max(0,\d v/\d r)$. The first approach estimates (approximately) an upper limit to the \CAK-like force in regions with negative velocity gradients, whereas the second approach would correspond to a lower limit. As most previous studies of line-driven dynamical flows have used 
$\left|{\d v}/{\d r}\right|$ 
(e.g., in investigations of line-driven winds from rapidly rotating stars, \citealt{Petrenz_00}, and around discs, \citealt{Proga_98,Kee_16}), we follow this standard approach in this paper. However, test computations using instead the lower limit $\max(0,\d v/\d r)$ show that  although some details of the radiation force profile then are changed, none of our conclusions are affected.}
\newtxt{With this approximation, using Eq. \ref{Eq_cak_g} for} \oldtxt{This} $\Gamma_\cak$ then accounts for the force from spectral lines in supersonic regions, while $\Gamma_\opal$ handles the force from the continuum and spectral lines in the static limit. As such, the total radiative acceleration can be simply estimated by the sum of the \CAK-like contribution and the static \opal opacities, $\Gamma_\tot = \Gamma_\opal + \Gamma_\cak$. 

Let us note here directly that the \cak line-force is originally derived for application in winds that are optically thin for continuum radiation. In the lower layers of the dense WR outflows considered here, this will typically not be the case. A more rigorous treatment would then need to take into account also the change in continuum intensity by solving the transfer equation for line+continuum opacities. While approximations for the diffusion limit exist \citep{Gayley_Owocki_95} no general formalism has been developed. As such, in this first study of WR wind dynamics we opt for the simple radial streaming \CAK-form above.\oldtxt{, keeping in mind the mentioned caveats.}

\newtxt{Furthermore, due to the explicit dependency of the \cak line-force on $\d v/\d r$ the sonic-point is now formally not a critical point (see \citealt{Castor_75}). However, our iteration scheme circumvents this issue by always applying the velocity gradient from the \textit{previous iteration}, meaning that $\Gamma_\tot(r)$ technically is no longer an explicit function of $\d v/\d r$. Nevertheless, since we require $\d v/\d r$ to be converged between iterations the corresponding feedback upon $\Gamma_\tot(r)$ is still (implicitly) accounted for. The agreement between the converged structures of these steady-state models and the time-dependent simulations presented in the next section brings further support to this method.}

A key question then becomes whether $\Gamma_\cak $ also has a significant impact on the  conditions at the optically thick lower boundary \newtxt{at the sonic point}. Assuming typical O-star values $\alpha = 0.66$, $\bar Q = 2300$ \citep[e.g.,][]{Puls_00}, approximating the velocity gradient\footnote{We define $v_{\max}$ as the maximum velocity found for the given fixed mass loss in the static opacity limit.} as $ \d v/\d r \sim \Delta v_{\max}/\Delta R(v=v_{\max})$, and using the mass density and velocity computed in the static opacity limit for $\dot M = 5.0\cdot 10^{-5}\UnitMd$, we find the Eddington ratio for the \cak  force at the lower boundary to be $\Gamma_\cak = g_\cak /g \sim 0.1$. Such a small contribution from the \cak  force is not surprising due to the high density at the lower boundary\footnote{The numerical simulations presented in \newtxt{Section} \oldtxt{Sect.} \ref{Se_steady_state}-\ref{Se_Time_dependant} \textit{a posteriori} confirm the simple order of magnitude estimate made here, showing explicitly that the force contribution from the CAK-like opacities near the lower boundary indeed is very small for our considered cases (see e.g. Fig. \ref{Fig_steady_gam}). This thus resolves some potential issues of our basic approach (simple summation of OPAL and CAK-like opacities) regarding 'double-counting' the effect of spectral lines in the static limit.} ($\Gamma_\cak \sim 1/\rho^\alpha$, with $\alpha > 0$). However, from test-calculations it turns out that although the \cak  contribution indeed increases as we move outwards in the wind, a simple model with constant $\alpha = 0.66$, $\bar Q = 2300$ is still not able to provide the force necessary to achieve a positive total wind energy and so drive the wind to infinity (see previous section). 

There are, however, several indications that the line-force in the outer winds of classical WR stars indeed might be significantly enhanced as compared to O-type stars. For example, the independent \newtxt{MC} \oldtxt{Monte-Carlo (MC)} line-force calculations by \citet{Lucy_93} and \citet{Springmann_98} both find global momentum rates for WR stars that exceed those of O-stars by an order of magnitude or more. In these MC computations, the outer-wind enhancement of the line-force stems from the decreasing level of ionisation with increasing distance from the star in the optically thick WR outflows. \newtxt{Due to this shift from high-to-low ionisation stages as photons move outwards in the wind, "new" spectral lines become available for photons to interact with. As these new lines have different interaction frequencies than those previously available in the inner wind parts, this means that the line-frequency gaps typically observed for the nearly constant ("frozen-in") ionisation conditions of optically thin O-star winds are effectively closed.}
\oldtxt{The highly stratified ionisation structure means that as photons are moving outwards in the wind, new fresh sets of lines from lower ionisation stages become available. Since these 
"new emerging" lines then have different interaction frequencies than those available in the lower wind parts, the line-frequency gaps typically observed for the nearly "frozen-in" ionisation conditions of optically thin O-star winds are effectively closed.} This then leads to "photon-trapping" \citep{Lucy_93} and efficient multi-line scattering, which significantly increases the line force.

\begin{figure}[ht]
\includegraphics[scale=0.19]{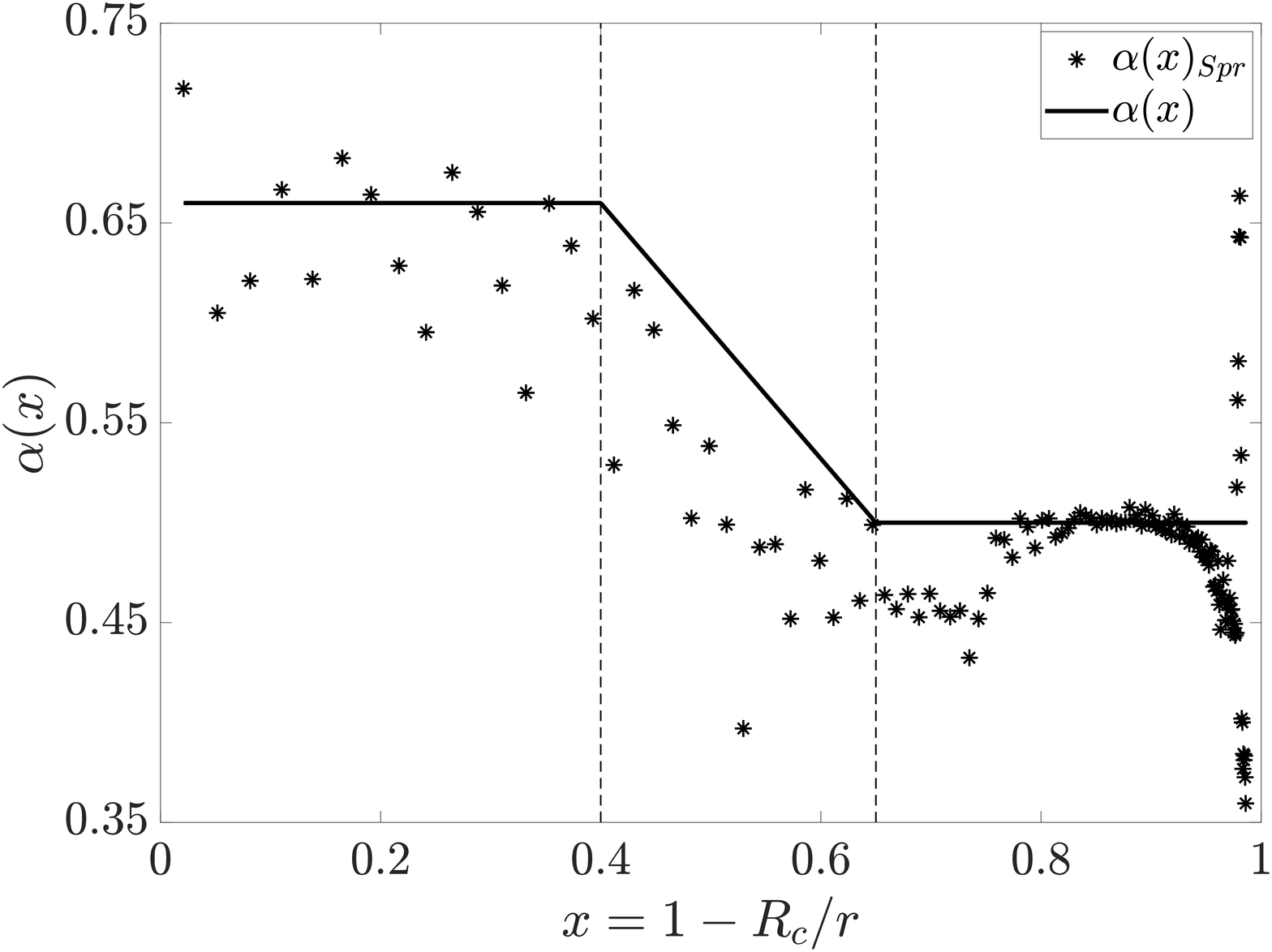}
\caption{This plot demonstrates values of $\alpha$ (see text) required to reproduce the line-force computed from MC radiative transfer calculations by \citet{Springmann_94} (asterisks). The solid line then over-plots the piecewise linear function adopted in this paper to describe the corresponding variation of $\alpha$. The vertical dashed lines correspond to $r_1$ ($x_1$) and $r_2$ ($x_2$).} 
\label{Fig_steady_Springman_alpha}
\end{figure}

To mimic the effect of an outer wind line-force enhancement, we introduce here a simple spatial variation $\alpha = \alpha(r)$, which increases the \cak force away from the star by a modest decrease of $\alpha$. To this end, we assume $\alpha(r)$ as a piecewise linear function with a constant $\alpha_{\max}$ from the lower bounder to $r_1$ and a constant $\alpha_{\rm min}$ 
from $r_2$ to the outer boundary, interpolating between the two values in the range $[r_1,r_2]$. 
This then allows the wind outflow initiated at the lower boundary to achieve positive total energy, and so be sustained until the upper boundary. While the specific treatment of $\alpha$ is done here in an ad-hoc manner, the overall assumed form is quite consistent with the multi-scattering line-force enhancement found in the MC line-force computations for a \CAK-like spectral line distribution by \citet{Springmann_94}. The approximate values of $\alpha(x)$ that correspond to the line-force computed by \citet{Springmann_94} are given in Fig. \ref{Fig_steady_Springman_alpha} as black asterisks, with the $\alpha(x)$ as adapted in the standard model of this paper over-plotted as a solid line. 
The figure shows that the order of magnitude outer-wind line-force enhancement applied here indeed is on the same order as that found in these MC calculations. 

We note, however, that the \citet{Springmann_94} calculations were performed only for parameterised (monotonic) $\beta$-type velocity laws. As such, there is not a complete one-to-one correspondence between the spatial and velocity scales in these MC calculations and those of the present paper. Nonetheless, the overall similarity with the very simple \newtxt{radial dependence of the $\alpha$ parameter} \oldtxt{$\alpha$-law} adopted here is encouraging. In a follow-up work, we plan to couple the hydrodynamic simulations presented here directly to such MC computations, building on the method developed by \citet{Puls_00} for proper estimation of occupation numbers, and accounting fully for multi-scattering effects as well as non-monotonic velocity fields.

As outlined in details in the following sections (for our standard values listed in Table \ref{Tab_cak_params}) the formalism outlined above gives rise to "dynamically inflated" wind solutions, where the structure in the optically thick deep layers near the iron bump is primarily governed by $\Gamma_\opal$ but where $\Gamma_\cak $ takes over the driving in the diluted outer regions. This then causes a typical wind structure that is characterised by a high mass-loss rate ignited from the lower boundary, a slow (end even partially negative) acceleration of the deeper layers, and a rapid re-acceleration of the outer wind. As such, the typical velocity laws we find not only drastically deviate from the standard $\beta$-type laws assumed by most 
atmospheric models, but they also contain a significant region of negative acceleration \citep[see also fig. 14 in][]{Pauldrach_93}. 

\begin{table}[ht]
    \centering
    \caption{Summary of standard line-force parameters used in this paper}
    \begin{tabular}{ c c c c c }
         \hline
         $\bar Q$ & $\alpha_{\rm max}$ & $\alpha_{\rm min}$ & $r_1/R_{\rm c}~(x_1)$ & $r_2/R_{\rm c}~(x_2)$\\
         \hline
         $2300$ & $0.66$ & $0.50$ & $1.7~(0.4)$ & $2.8~(0.65)$ \\
         \hline
    \end{tabular}
    \label{Tab_cak_params}
\end{table}

\subsection{Dynamically inflated steady-state model}\label{Se_steady_state_dynamic_inflation} 

Based on the "hybrid" formalism introduced above we can now numerically solve Eqs. \ref{Eq_Lucy}-\ref{Eq_motion_simp}, iterating toward convergence using 
a simple scheme based on Runge-Kutta integrations. 
As discussed in Section \ref{Se_steady_state_twopointer_boundary}, to constrain the mass-loss rate it is critical to here take into account the two-pointed boundary value nature of the problem. 
At a basic level, every iteration is performed using two steps: i) an inside-out integration $R_{\rm c}\rightarrow R_{\rm max}$, where the velocity, density, and temperature structure are computed, followed by ii) an outside-in integration $R_{\rm max}\rightarrow R_{\rm c}$ updating the temperature structure using the correct boundary condition from the optical depth found in the first step. The mass-loss rate is then updated for the next iteration by finding the correct location on the 
$\Gamma = 1$ curve at the lower boundary, using the updated temperature structure from the outside-in integration.

\begin{figure*}[ht]
\centering
    \begin{subfigure}[t]{.34\textwidth}
    \includegraphics[width=\textwidth]{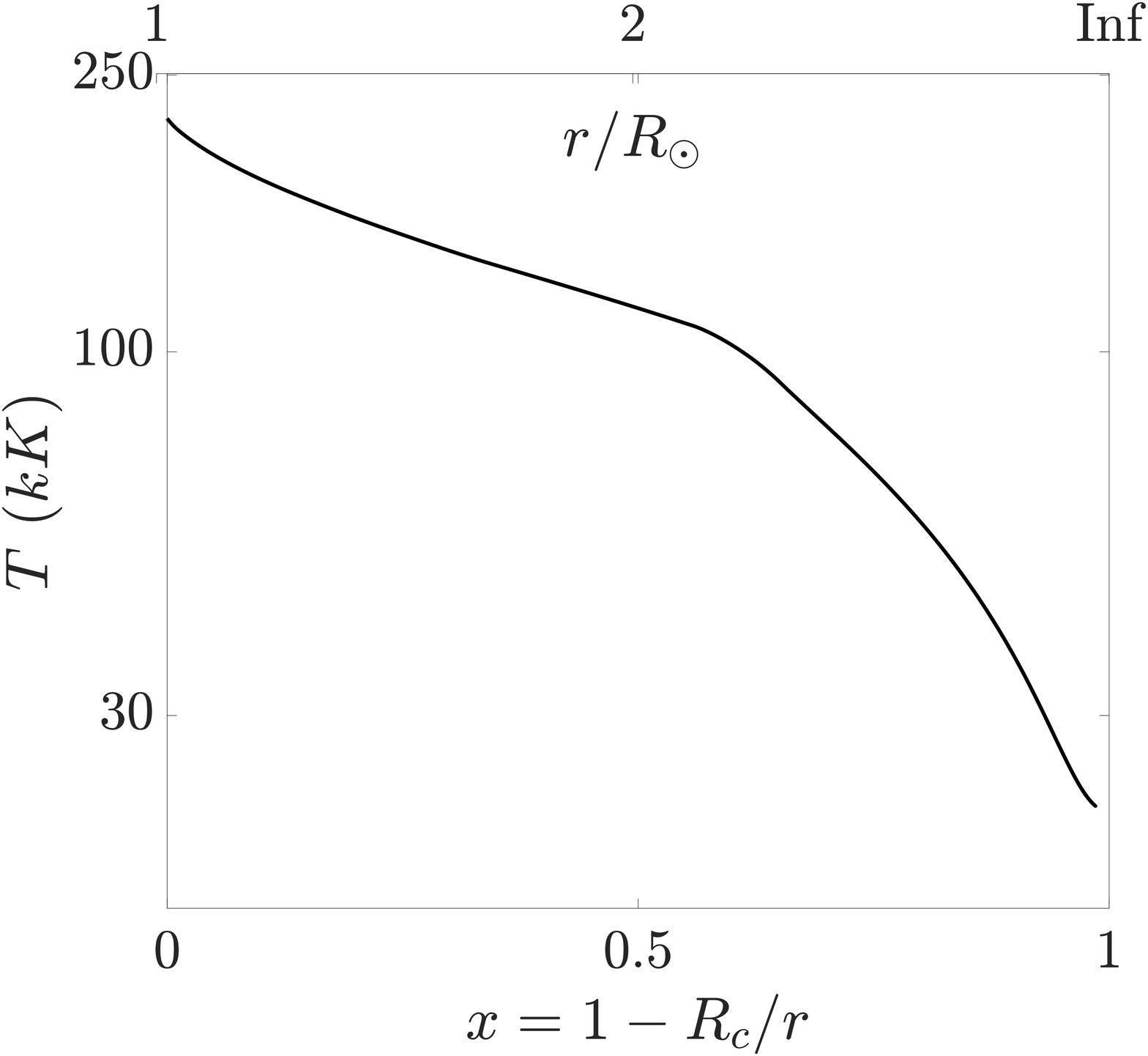}
    \caption{}
    \label{Fig_steady_temp}
    \end{subfigure}%
    \begin{subfigure}[t]{0.34\textwidth}
    \includegraphics[width=\textwidth]{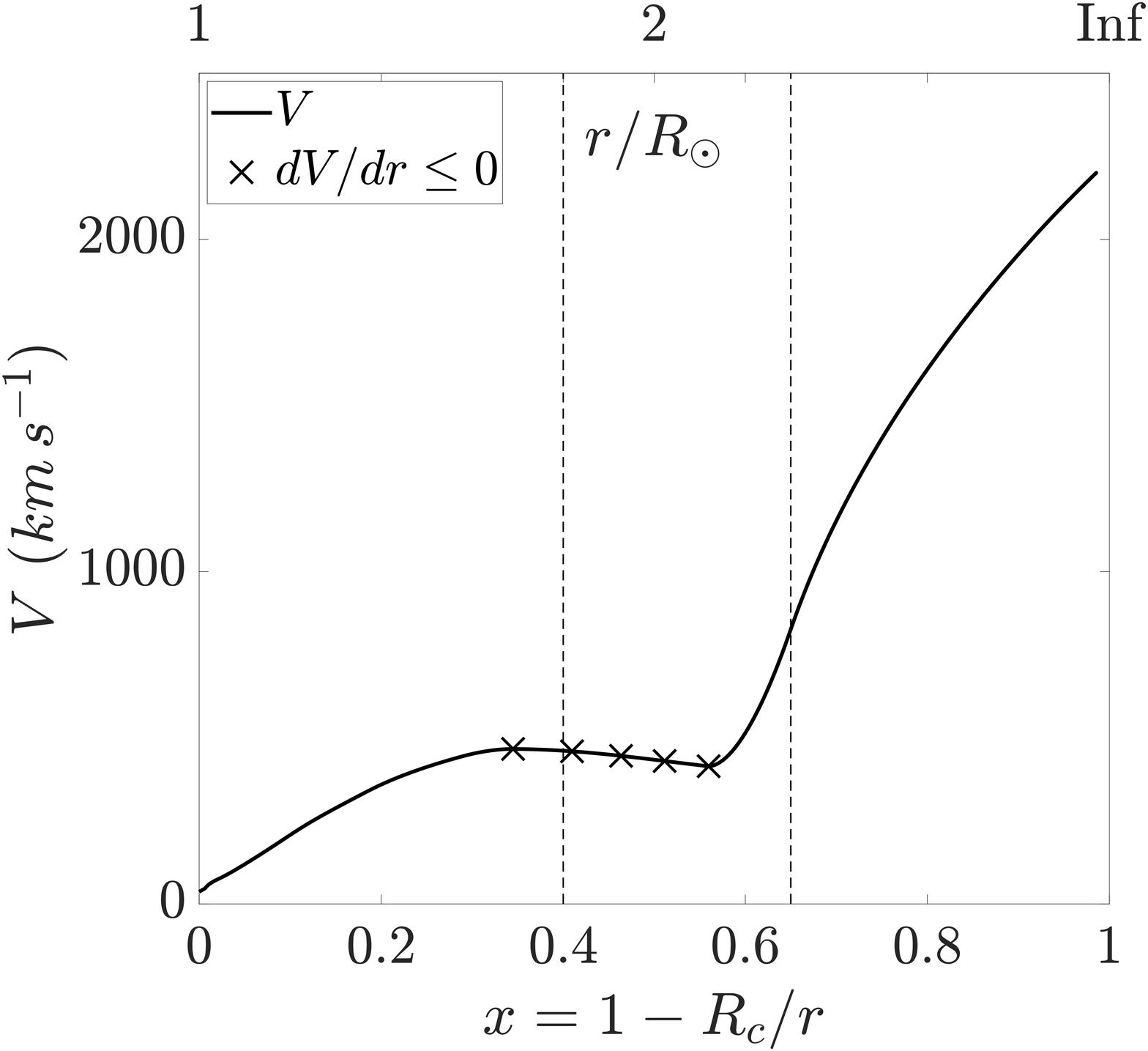}
    \caption{} 
    \label{Fig_steady_velo}
    \end{subfigure}%
    \begin{subfigure}[t]{0.34\textwidth}
    \includegraphics[width=\textwidth]{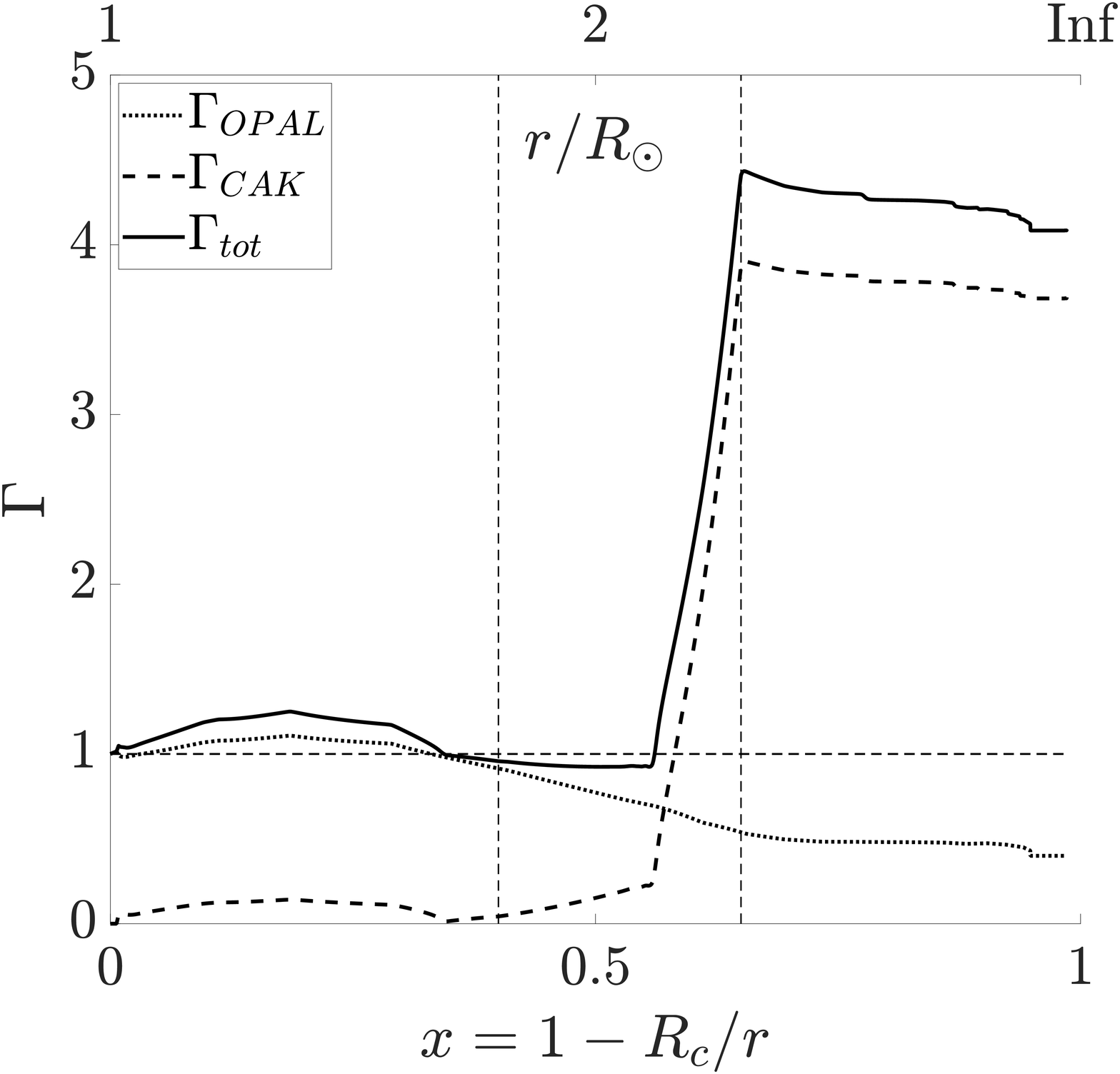}
    \caption{} 
    \label{Fig_steady_gam}
    \end{subfigure}\\%
    \begin{subfigure}[t]{0.33\textwidth}
    \includegraphics[width=\textwidth]{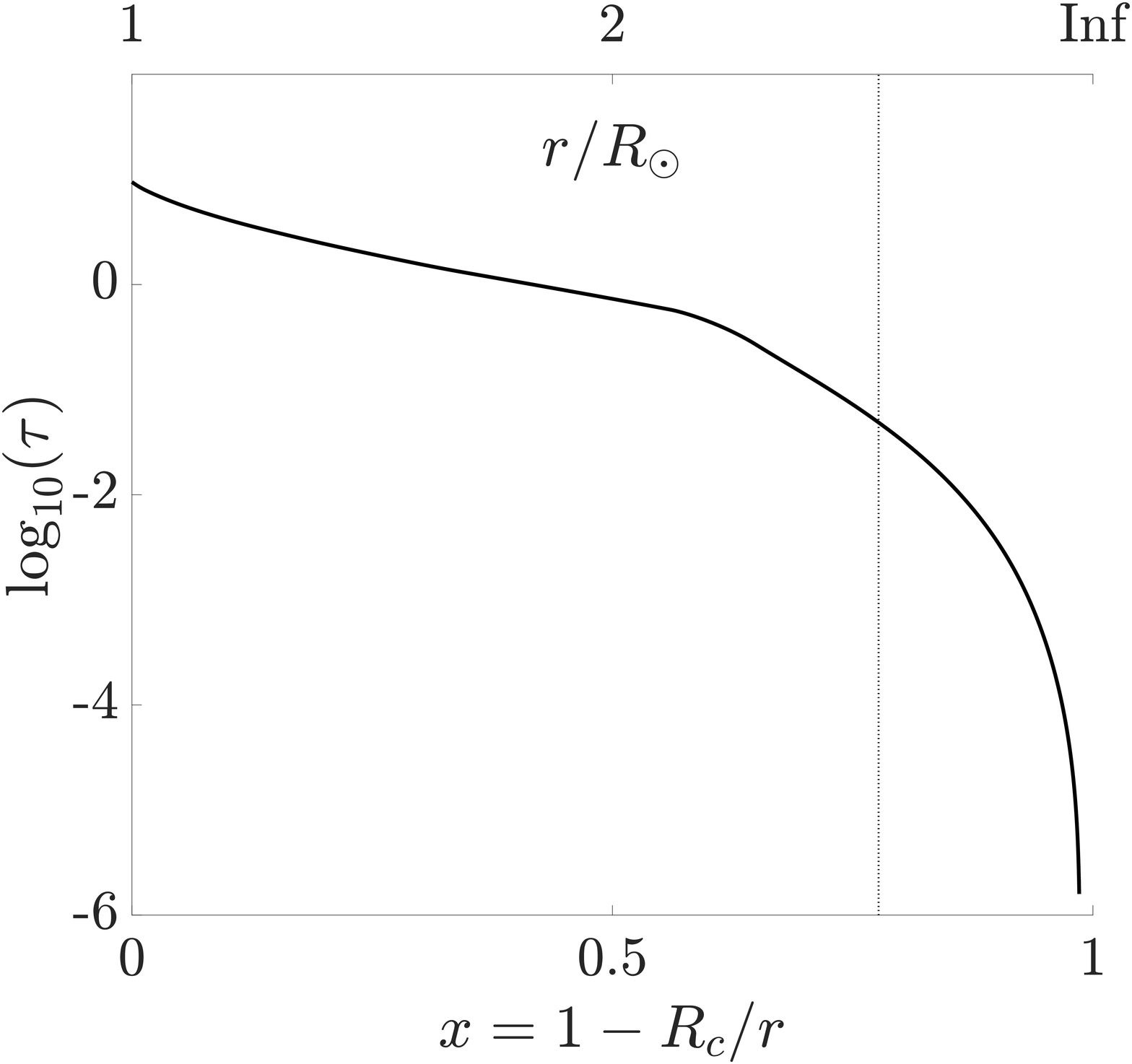}
    \caption{}
    \label{Fig_steady_opt_depth}
    \end{subfigure}%
    \begin{subfigure}[t]{.33\textwidth}
    \includegraphics[width=\textwidth]{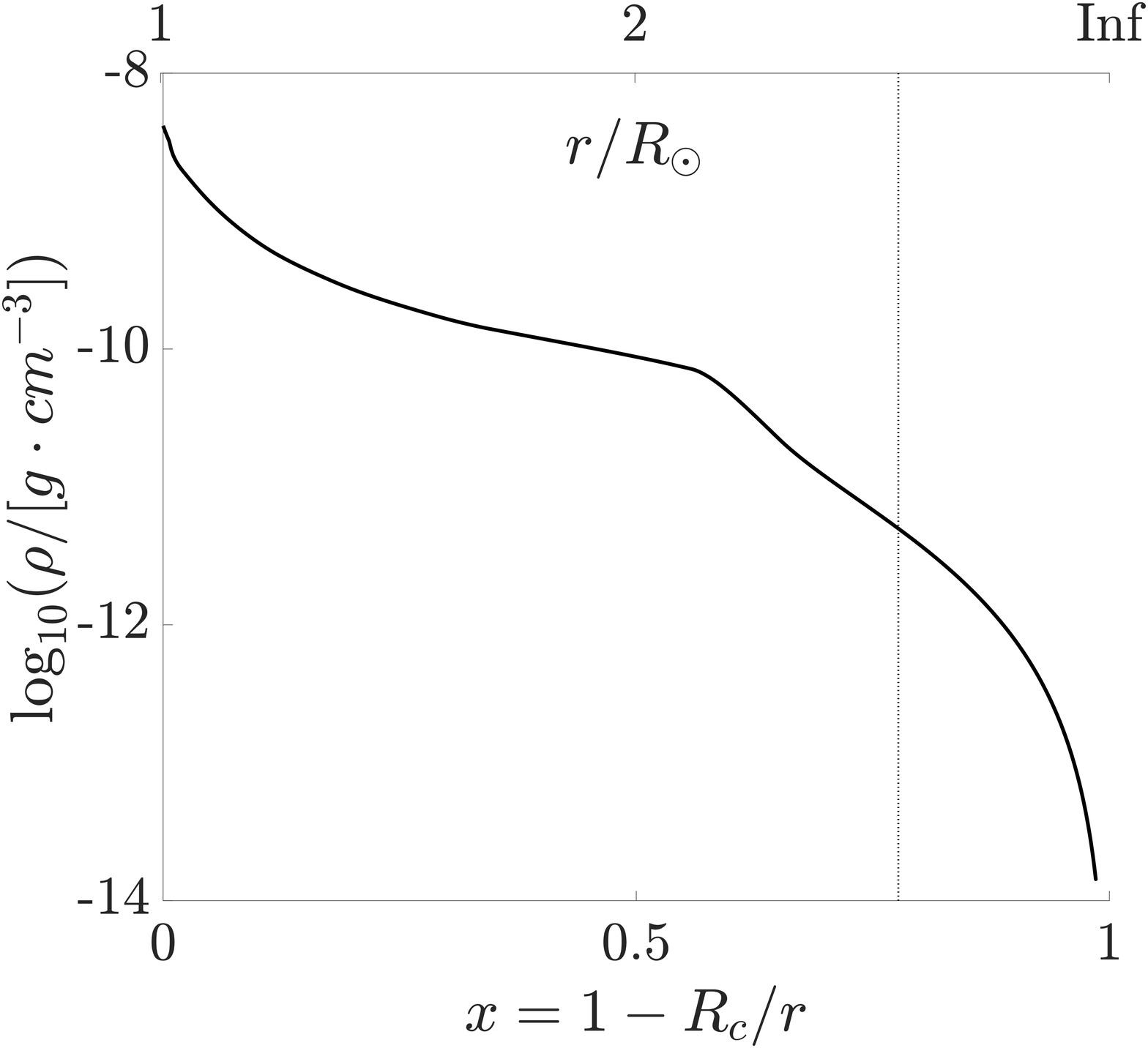}
    \caption{}
    \label{Fig_steady_dens}
    \end{subfigure}
    \caption{ Solutions of a steady-state model with stellar parameters as in Table \ref{Tab_cak_params}. \textit{(a)} temperature profile. \textit{(b)} Velocity profile in solid line, the region of stagnation or negative acceleration in {\sffamily X} markers. \textit{(c)} Eddington ratios for full radiation force in solid line, only the \opal contribution in dotted line, only \cak contribution in dashed line, and  $\Gamma = 1$ in thin horizontal dashed line. \textit{(d)} Profile of spherically modified optical depth. \textit{(e)} Density Profile. On panels (d,e) the vertical dotted line at \newtxt{$R_{\rm ph} = 4.48R_{\rm c} = 4.48R_\odot$} \oldtxt{$R_{\rm ph} = 4.48R_\odot$} shows the location of the photosphere. \newtxt{The top axis gives the radial distance in units of $R_\odot$, but since in this paper $R_{\rm c} = R_\odot$ this is also equivalent to normalising the radius to $R_{\rm c}$.}}
\end{figure*}

Using this scheme, Figure \ref{Fig_steady_temp} shows  the final, converged temperature profile for our standard stellar parameters (Table \ref{Tab_stellar_params}) and choice of line-force parameters (Table \ref{Tab_cak_params}). The final (also converged) mass-loss rate for the model is $\dot{M} = 1.47\cdot 10^{-5}\UnitMd$.

In addition to a high predicted mass-loss rate, the solution displays a non-monotonic velocity field. The resulting velocity profile  shown in Fig. \ref{Fig_steady_velo} can be \oldtxt{basically} split into three \newtxt{basic} regions: {\it 1)} an initial acceleration from the core followed by {\it 2)} stagnation and even a deceleration part (and so a negative velocity gradient, marked with crosses in the figure), and {\it 3)} an outer region of fast re-acceleration to outflow velocities $> 1000\UnitV$. The distinct behaviour of these three regions can be understood via the different contributing parts of the total radiation force.

Figure \ref{Fig_steady_gam} illustrates this, plotting the individual contributions of \opal and \CAK-like opacities to the total radiative acceleration. From the figure, on one hand,  it is clear that in the inner regions the radiation force is dominated by \opal opacities and the \CAK-like contribution is small; indeed, the sonic point conditions at the lower boundary can here be quite well approximated using only \opal opacities (i.e., neglecting any influence from the Doppler-shift enhancement of line-opacities). On the other hand, in the outer regions $\Gamma_\cak $ becomes the dominant part, causing a re-acceleration of the flow and setting a high terminal wind velocity\newtxt{. This re-acceleration} \oldtxt{and that} ensures that the energy requirement of Eq. \ref{Eq_net_en} is satisfied. As such, it is this part that ultimately allows the wind to be sustained all the way to the outer boundary, in contrast to the "failed outflow" and fallback seen in the previous section. 

To explore the influence of different opacity contributions on the mass loss of the model, we explore the logarithm of the spherically modified optical depth (Eq. \ref{Eq_tau_sp}) as a function of radius in Fig. \ref{Fig_steady_opt_depth}. The dashed vertical line marks the location of the stellar photosphere defined by  Eq.~\ref{Eq_Tpheff}. (Note, however, that at the photosphere $\tau_{\rm sp}(R_{\rm ph}) \neq 2/3$ as $R_{\rm c}$ is used as a scaling radius in Eq. \ref{Eq_tau_sp}.) For most stars, this photospheric radius $R_{\rm ph}$ serves as the hydrostatic stellar radius; by contrast, for the WR stars examined in this paper, the photosphere is located well in the outflowing regions. Inspection of Fig. \ref{Fig_steady_opt_depth} shows that the majority of $\tau_{\rm sp}$ is accumulated within few stellar radii in the regions dominated by \OPAL-opacities, thus the converged lower boundary temperature (and thus the mass-loss rate) is primarily controlled by $\Gamma_\opal$.

Inspection of the density distribution in Fig. \ref{Fig_steady_dens} further clarifies the terminology "dynamic inflation". In this figure, we have again identified the location $R_{\rm ph}$ with a vertical dashed line. The figure shows a relatively slow decline of the density in the sub-photospheric parts, which is somewhat reminiscent of the profiles seen in hydrostatic stellar models near the Eddington limit that undergo envelope inflation (see further below). 
But unlike such static inflation models, the simulations here allow for a velocity field to develop; the low-density envelope now represents a "dynamically inflated" star. 

\subsection{Comparison to static inflation}\label{Se_steady_state_static_inflation}

For the same lower boundary conditions as adopted above, we can also compute static models. This can then serve as a direct comparison between the dynamic inflation found in the models above, and the static inflation that often occurs in stellar structure models neglecting the $v \, \d v/\d r$ term in the e.o.m. (and so assuming a hydrostatic stellar envelope throughout, e.g., \citealt{Petrovic_06,Grafener_12}).

To construct such static models, we simply set $v\sim \d v/\d r\sim 0$ in our steady-state e.o.m. above. Introducing a column mass
\begin{equation} 
    \frac{\d r}{\d m} = \frac{1}{\rho}\,,
\end{equation} 
we write the resulting equation of hydrostatic equilibrium as:
\begin{equation}\label{Eq_static_pgas}
    \frac{\d P_{\rm g}}{\d m} = \frac{M_\star G}{r^2}(\Gamma - 1)\,. 
\end{equation}
We further simplify the temperature calculation by assuming radiative diffusion in these optically thick hydrostatic layers: 
\begin{equation}\label{Eq_static_prad}
    \frac{\d P_{\rm r}}{\d m} = -\frac{M_\star G}{r^2}\Gamma\,,
\end{equation} 
for a radiation pressure according to Eq. \ref{Eq_rad_pressure}. Combining the equations for gas and radiation pressure, one obtains 
\begin{equation} 
    \frac{\d P_{\rm g}}{\d P_{\rm r}} = \frac{1 - \Gamma}{\Gamma}\,,  
\end{equation} 
where now $\Gamma = \Gamma_\opal$ throughout the complete model. 

This set-up imposes a similar two-point boundary value problem as previously analysed. To provide a fair comparison with those dynamical simulations, we assume the same lower boundary radius $r = R_\odot$ as before and re-use the lower boundary temperature 
$T \approx 230\UnitT$ found from the converged hydrodynamic steady-state model presented above. However, at this inner boundary we now impose $\Gamma < 1$ as $\Gamma\geq 1$ would lead to non-zero gas pressure at $r \rightarrow\infty$ \citep[see also Section 2.2 in][]{Grafener_12}. 
The outer boundary conditions for gas and radiation pressure (i.e., density and temperature) are also set up in the same way as before. Assuming $\Gamma = const = \Gamma_e$ for the region $r\in[R_{\rm max},~ \infty)$, we may integrate the hydrostatic equations to estimate 
the temperature and the density at the outer boundary $r = R_{\rm max}$. 
This fixes the "effective temperature" and the spherically modified optical depth scale $\d \tilde \tau_{\rm sp} = - \kappa_\opal \,\rho \,(R_{\max}/r)^2 \,\d r$ of these hydrostatic (HS) models, requiring that the outer boundary always satisfies $\tilde\tau_{\rm sp}(R_{\max}) = 4/3$. Thus in this model, the "stellar photospheric radius" is defined as
\begin{equation} 
    \tilde R_{\rm ph} \equiv r(\tilde\tau_{\rm sp} = 4/3), 
\end{equation} 
for the corresponding "hydrostatic effective temperature" 
\begin{equation} 
   T_{\rm HS,\, eff}^4 \equiv \frac{\L}{4 \pi \sigma \tilde R_{\rm ph}^2}\,, 
\end{equation}
quite analogous to how these quantities typically are defined in static stellar structure and evolution models.

\begin{figure*}
\centering
  \begin{subfigure}[t]{.48\textwidth}
    \centering
    \includegraphics[width=\textwidth]{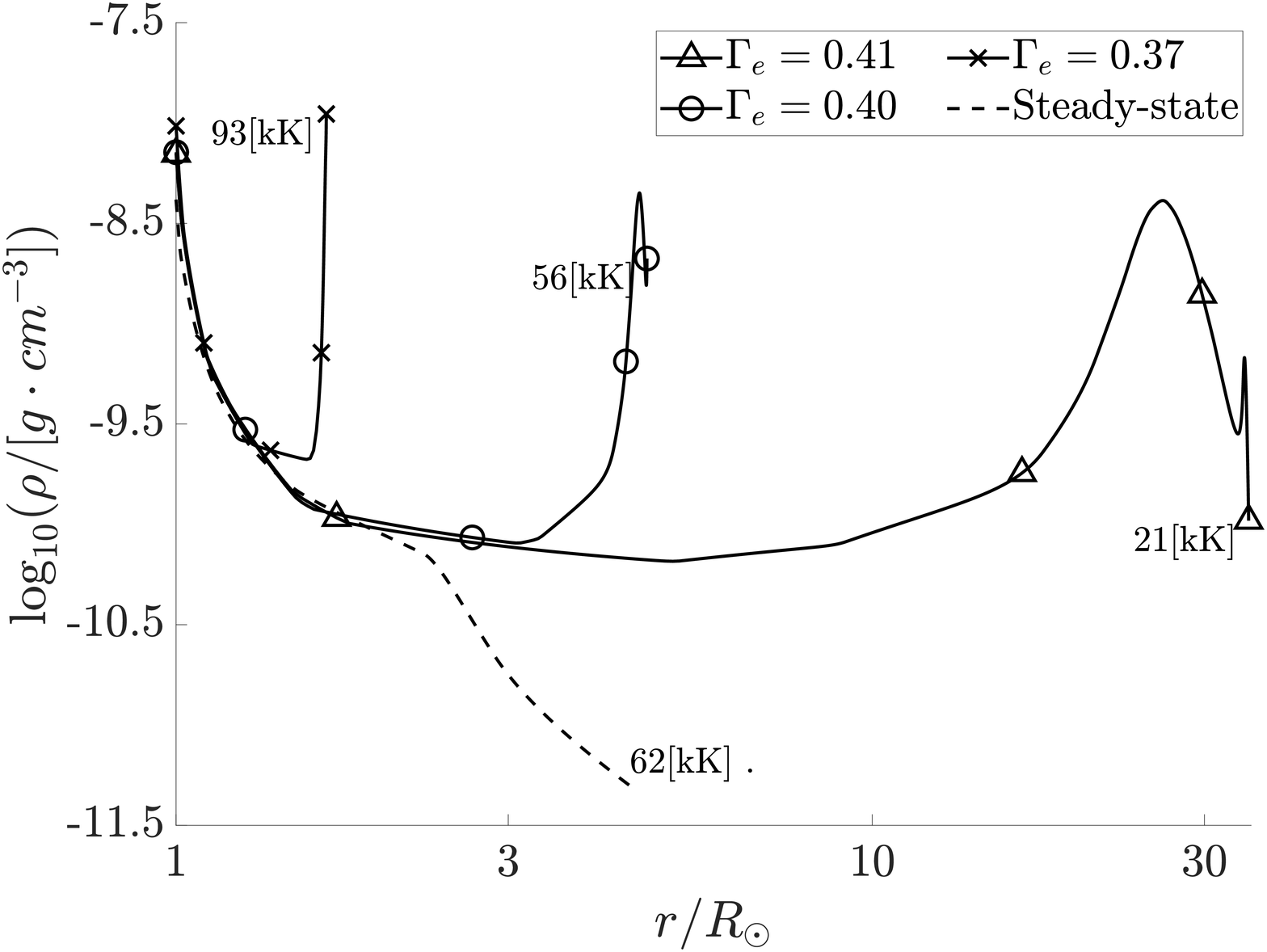}
  \end{subfigure}%
  \hfill%
  \begin{subfigure}[t]{.48\linewidth}
    \centering
    \includegraphics[width=\textwidth]{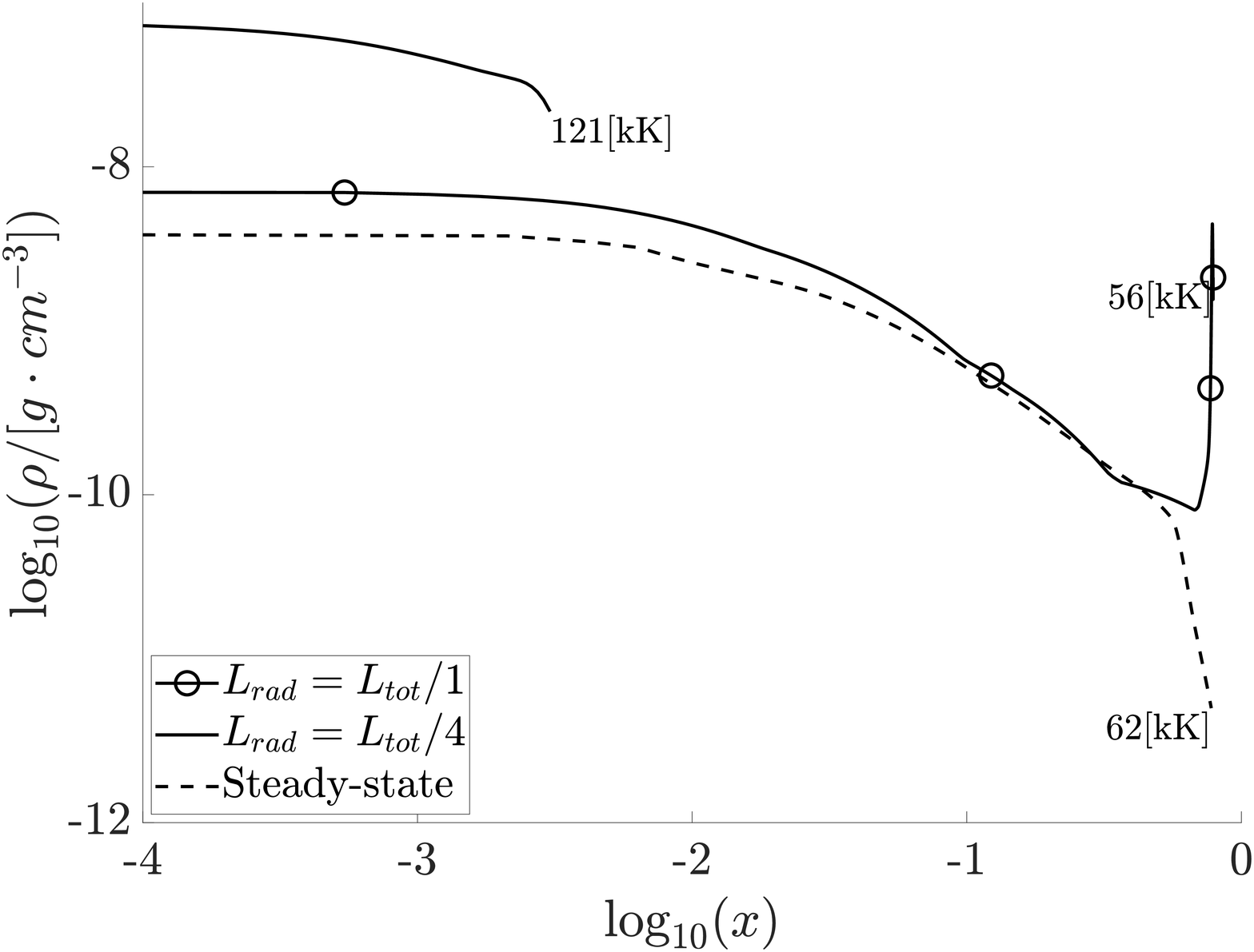}
  \end{subfigure}
\caption{Density profiles of different static and steady-state solutions. On both plots the solid lines show the different cases of hydrostatic solutions and dashed lines show the sub-photospheric part of the steady-state outflow (see Fig. \ref{Fig_steady_dens}). The different markers then compare hydrostatic solutions for different assumed Thompson scattering Eddington factor $\Gamma_e$ (left panel) and for different assumed $\L/L_{\rm rad}$ ratios for the fixed total luminosity $\log_{10}(\L/\L_\odot) = 5.416$ (right panel). The numbers at the right end of each density profile correspond to the "photospheric effective temperature" of the corresponding model in kilo-Kelvins, as defined in the text. To better visualise the innermost parts of the simulations, the right panel displays the logarithm of the scaled radius $x=1 - R_{\rm c}/r$ on the abscissa.}
\label{Fig_static_density}
\end{figure*}

These boundary conditions are used to numerically solve the new set of equations from the outer boundary inwards. A consistent solution is obtained by shooting for the fixed inner boundary temperature and radius, varying the stellar photospheric radius $\tilde R_{\rm ph}$ and assuming inner-boundary values as discussed above. In this way we obtain static envelopes corresponding directly to our previous dynamic models. The right panel of \newtxt{Fig.}\oldtxt{Figure} \ref{Fig_static_density} shows the predicted densities for static models with different assumed\footnote{For the  fixed stellar mass and electron scattering mass absorption coefficient the variation of $\Gamma_e$ is equivalent to the variation of $L_{\rm rad}$} $\Gamma_e\in[0.37,\;0.40,\; 0.41]$ \citep[for comparison see Fig.~1 in][]{Grafener_12}. These densities are then compared to those of the steady-state computation in Section \ref{Se_steady_state}, for which $\Gamma_e = 0.40$. %

We note directly that in the static case an extended region of low density is formed. The photospheric radius $\tilde R_{\rm ph}$ of this "inflated envelope" is extremely sensitive to $\Gamma_e$, \newtxt{with $\tilde R_{\rm ph}$ growing notably as $\Gamma_e$ is increased} \oldtxt{becoming larger with increasing values}. For the static model with highest assumed $\Gamma_e = 0.41$, we already find a large $\tilde R_{\rm ph} \approx 35 R_\odot$ accompanied by a relatively cool effective temperature $T_{\rm HS,\, eff} = 21\UnitT$; further increasing $\Gamma_{e}$ causes the inferred photospheric radius to inflate to increasingly unrealistic, extremely large values. 
For the $\Gamma_e = 0.40$ that we use in our dynamic models we find $\tilde R_{\rm ph} = 4.7 R_\odot$ and $T_{\rm HS,\,eff} = 56\UnitT$. The reason for the inflation here is the same as in the models by \citet{Petrovic_06} and \citet{Grafener_12}. Basically, it is related to the fact that for high values of $\Gamma_e$ the integration moves along a $\Gamma \approx 1$ curve, forcing the density to remain low and almost constant over a large radial extent. Since enough column mass (or equivalently optical depth) still has to be accumulated in order to meet the lower boundary condition, the only way the star can react is by expanding in radius \citep[see also discussion in][]{Owocki_14}. Comparing this now to our dynamic solutions shows that, for the same $\Gamma_e$, inclusion of the dynamical terms reduces the inflation somewhat. Nonetheless, $R_{\rm ph}$ still lies very far away from the core in such outflow models of dynamically inflated envelopes; indeed, for our standard $\Gamma_e = 0.40$ we obtain $R_{\rm ph} = 4.48 R_\odot$ and $T_{\rm ph,\,eff} = 62\UnitT$. 
Such a close agreement of photospheric radii in dynamic and static models is however not generally found. Namely, a very small increase to $\Gamma_e=0.41$ in the static model leads to an order of magnitude increase of the star's photospheric radius (see \newtxt{Fig} \oldtxt{fig}. \ref{Fig_static_density}), in contrast to the dynamic model where this barely affects the location of $R_{\rm ph}$.
 
The analysis above assumes that all luminosity is carried by radiation. In stellar models, the strong increase in opacity around the iron-bump will typically lead to an onset of convection, such that $\L = L_{\rm rad} + L_{\rm conv}$. In principle, this may alleviate the inflation by allowing a portion of the luminosity to be carried by convection, lowering the effective $\Gamma_e \sim L_{\rm rad}$. However, for the conditions prevailing in WR stars convection in these near-surface layers should be very inefficient. Following \citet{Grafener_12}, an upper limit to the fraction of the total stellar flux $F$ that can be carried by convection is:  
 \begin{equation} 
    \frac{F_{\rm conv}}{F}
    \approx 0.0034 \frac{10^5\,\mathrm{K} }{T_{\rm c,\, eff}}\,.
\end{equation} 
This means that for the $T_{\rm c,\, eff} > 10^5\UnitT$ considered in this paper, less than a percent of the total luminosity will be carried by convection.

This convective inefficiency is confirmed by the computation of WR stars using the stellar structure and evolution code package MESA \citep{Paxton_19}. Namely, using the standard mixing-length theory (MLT) prescription for convection shows that such stellar structure and evolution computations\footnote{Detailed description of these models as well as Input files to reproduce our MESA simulations are provided at \url{https://doi.org/10.5281/zenodo.4054811}} indeed undergo similar inflation in the WR stage as seen here in Fig. \ref{Fig_static_density}. However, in these stellar evolution simulations envelope inflation and the associated density inversions often imply prohibitively short time-steps within the model-calculations. To prevent this from happening, such stellar models often invoke an additional energy transport that is far greater than that implied by the upper-limit estimate of $L_{\rm conv}$ above \citep[see, for example, the so-called MLT++ prescription in MESA, Section 7.2 in][]{Paxton_13}. This then forces $L_{\rm rad}$ to always be low enough such that the calculation can proceed.

To mimic these "tricks", in the next step we artificially reduce $\Gamma_e\sim L_{rad}$ by introducing $\L = L_{\rm conv} + L_{\rm rad}$. Then using $L_{\rm rad}$ in Eq. \ref{Eq_static_pgas}, while computing the radiation pressure from the total luminosity in Eq. \ref{Eq_static_prad}, we solve the differential equations exactly as above. By then assuming different ratios $\L/L_{\rm rad}$ for the same fixed total luminosity, we can directly compare dynamic and static inflation models with and without such a reduction in radiative luminosity. Analogous to the left panel, the right panel of Fig. \ref{Fig_static_density} shows the radial density structure of a case with reduced radiative luminosity $\L/L_{\rm rad} = 4$ (solid black line), comparing this to the original case $\L/L_{\rm rad} = 1$  (solid-marked line). The ratio of $\L/L_{\rm rad} = 4$ here is chosen such that the stellar envelope no longer experiences a density inversion. The figure demonstrates clearly that the two solutions have vastly different radial scales; while the original model had $\tilde R_{\rm ph} = 4.7 R_\odot$ and $T_{\rm HS,\, eff} = 56\UnitT$, the model with reduced radiative luminosity has a very hot $T_{\rm HS,\, eff} = 121\UnitT$ and a photospheric radius that lies less than a percent above the core $R_{\rm c} = R_\odot$. 
 
The reason for these drastically different stellar photospheric radii is the same as described earlier; including an additional (artificial) energy transport reduces the effective $\Gamma_e\sim L_{\rm rad}$, so that a dense exponential atmosphere with small scale-height is formed. This then avoids inflation and density inversions, and the upshot is a hot and compact WR stellar surface. This is a good example of how envelope inflation and density inversions are often avoided in stellar structure computations by reducing $L_{\rm rad}$ well beyond the limits implied by standard convective energy transport.

A key result of our analysis here is thus that such "tricks" of the near-surface regions in WR stars are necessary only because of enforcing a hydrostatic solution. As shown above, when including the dynamical terms the stellar envelope will instead quite naturally develop a "dynamically inflated" outflow region. 

\section{Time-dependent numerical hydrodynamic simulations}\label{Se_Time_dependant}

The previous section presented calculations in the steady-state limit, using various degrees of approximations for solving the dynamical (and static) equations. Building on the same basic formalism, this section now presents full time-dependent numerical hydrodynamics simulations for spherically symmetric WR wind outflows. A key objective is to examine whether such time-dependent simulations relax to similar solutions as the simplified steady models of Section \ref{Se_steady_state}. 

The hydrodynamic simulations are performed using the finite volume code \texttt{MPI-AMRVAC} \citep{Xia_18} to solve \newtxt{Eqs.} \oldtxt{equations} \ref{Eq_time_continuity}, \ref{Eq_time_momentum}, and \ref{Eq_Lucy}. We use the 
\texttt{HLL} solver \citep{Harten_83} with a \texttt{MINMOD} flux limiter \citep{Van_Leer_79}.
The radiative acceleration is computed as in Section \ref{Se_steady_state_faild_wind} by summing up the contributions from \opal and \CAK-like opacities, 
$\Gamma_\tot = \Gamma_\opal + \Gamma_\cak $. Here the \opal contribution is trivially obtained from the local density and temperature, and the \cak  contribution from a simple central finite difference scheme applied to Eq. \ref{Eq_cak_g}. 
To avoid information propagating over multiple cells in a single integration 
we further limit the integration time-step $\d t$ by\newtxt{:} 
\[\d t = \min\left(\d\, t_{\rm CFL},\,0.3\sqrt{\frac{\Delta r}{g + g_r}}\right)\,\newtxt{,}\]
where $\d t_{\rm CFL}$ is the time-step set by the standard Courant-Friedrichs-Lewy condition 
\citep{Courant_28}, and $\Delta r$ the width of one numerical cell.
 
As Eq. \ref{Eq_Lucy}  is not the standard way of computing the energy balance in 
\texttt{MPI-AMRVAC} we have implemented an additional routine to handle this. In this subroutine the optical depth Eq. \ref{Eq_tau_sp} is first computed by 
trapezoid integration starting from the outer boundary $i = i_{\rm max}$, setting $\tau(i_{\rm max}) = \tau_{\rm out}$ as before, inwards to $i =i_{\rm min}$.
The radial temperature structure $T_i$ is then updated at each hydrodynamical time step according to Eq. \ref{Eq_Lucy}.
For simplicity, we here define $R_{\rm c}$ to always be at the lower boundary of the simulation.

\subsection{Initial and boundary conditions}\label{Se_Time_dependant_boundaries}

Analogous to our steady-state simulations, we define the outer boundary to be located at some radius $R_{\rm max}$ and use the same optical depth and temperature conditions as described in the previous section, Eq. \ref{Eq_Lucy} and the pre-specified minimum temperature (see above). As in e.g., \citet{Driessen_19}, the numerical outer boundary conditions on the velocity and density are set by simple extrapolation from the outer most points of the simulation grid. To properly resolve the sonic-point we set the inner boundary to a lower radius than the expected sonic-point $R_{\rm a}$, i.e. now $R_{\rm \min} \equiv R_{\rm c} = 0.99R_\odot < R_{\rm a}$. At this $R_{\rm \min}$ we then fix the lower boundary density $\rho_\star$ and allow the velocity to adjust to the overlying wind conditions \citep[see, e.g., ][]{Owocki_88}. $\rho_\star$ is initially estimated from our corresponding steady-state models by simply assuming hydrostatic equilibrium below the sonic-point.  The stellar mass and luminosity still are as in \newtxt{Table} \oldtxt{Tab.} \ref{Tab_stellar_params}.

The initial wind condition is an analytic outflow structure, in the form of a simple "$\beta$-law": 
\begin{equation}\label{Eq_beta_velocity}
v_\beta = v_\infty\left(1 - b\frac{R_{\rm c}}{r}\right)^{\beta}\,,
\end{equation}
where the terminal velocity $v_\infty^2 = 2 M_\star G(\Gamma- 1)/R_{\rm c}$, and $\beta = 0.5$ are derived from a simple assumed $\Gamma  = const > 1$. The  parameter $b$ sets the minimum initial velocity $\varv_{\rm min}$ at $r = R_{\rm c}$, here chosen such that $v_{\rm min} \approx 50\UnitV$. The equation of continuity then directly provides the initial density structure. Finally, the initial temperature structure is computed from Eq. \ref{Eq_Lucy} with the opacity set according to the assumed input $\Gamma$ \newtxt{from} above. Using these initial and boundary conditions, numerical time integration is carried out until the wind relaxes to a (quasi) steady state. We, however, note that the results of our simulations are not sensitive to these assumed initial conditions.

\subsection{Time dependent dynamic inflation model}\label{Se_Time_dependant_model}

\begin{figure}[ht]
\centering
\includegraphics[scale=0.19]{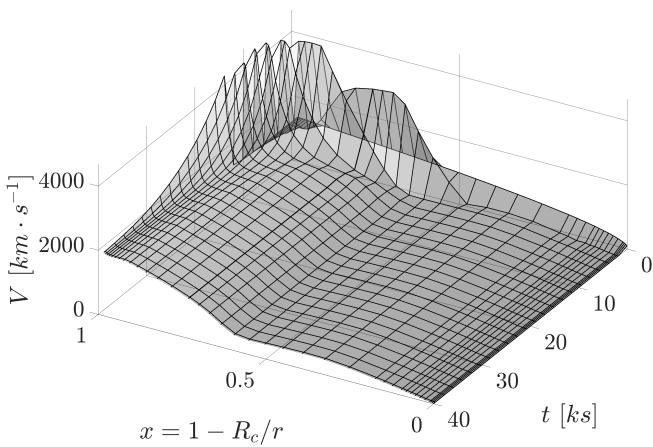}
\caption{Surface plot of velocity as a function of time and reduced radius $x$ for the model as described in text. The figure demonstrates how the velocity profile relaxes from the initially assumed $v_\beta$ to (quasi) steady-state profile over $\approx 40\Unittime$. }
\label{Fig_Time_dependent_velo_relax}
\end{figure}

Starting from the initially assumed structure the wind outflow gradually relaxes to a hydrodynamically consistent structure. The velocity-relaxation is displayed in Fig. \ref{Fig_Time_dependent_velo_relax} as a surface plot, where along one horizontal axis the physical time $t\Unittime$ of the simulation is shown and along the other the radius in $x = 1 - R_{\rm c}/r$. The vertical axis then measures velocity (in $\UnitV$). At $t=0$ the initial velocity structure, $v_\beta$, is visible, however as time increases this $\beta$-type structure gradually evolves toward the three-phased velocity profile discussed in \newtxt{Section} \oldtxt{Sect.} \ref{Se_steady_state_dynamic_inflation}. Around  at $t \approx 40\Unittime$ the simulation has relaxed to a state very similar to that found in our corresponding steady-state model (Section~\ref{Se_steady_state_dynamic_inflation}).

\begin{figure*}[ht]
\centering
  \begin{subfigure}[t]{.48\textwidth}
    \centering
    \includegraphics[width=\textwidth]{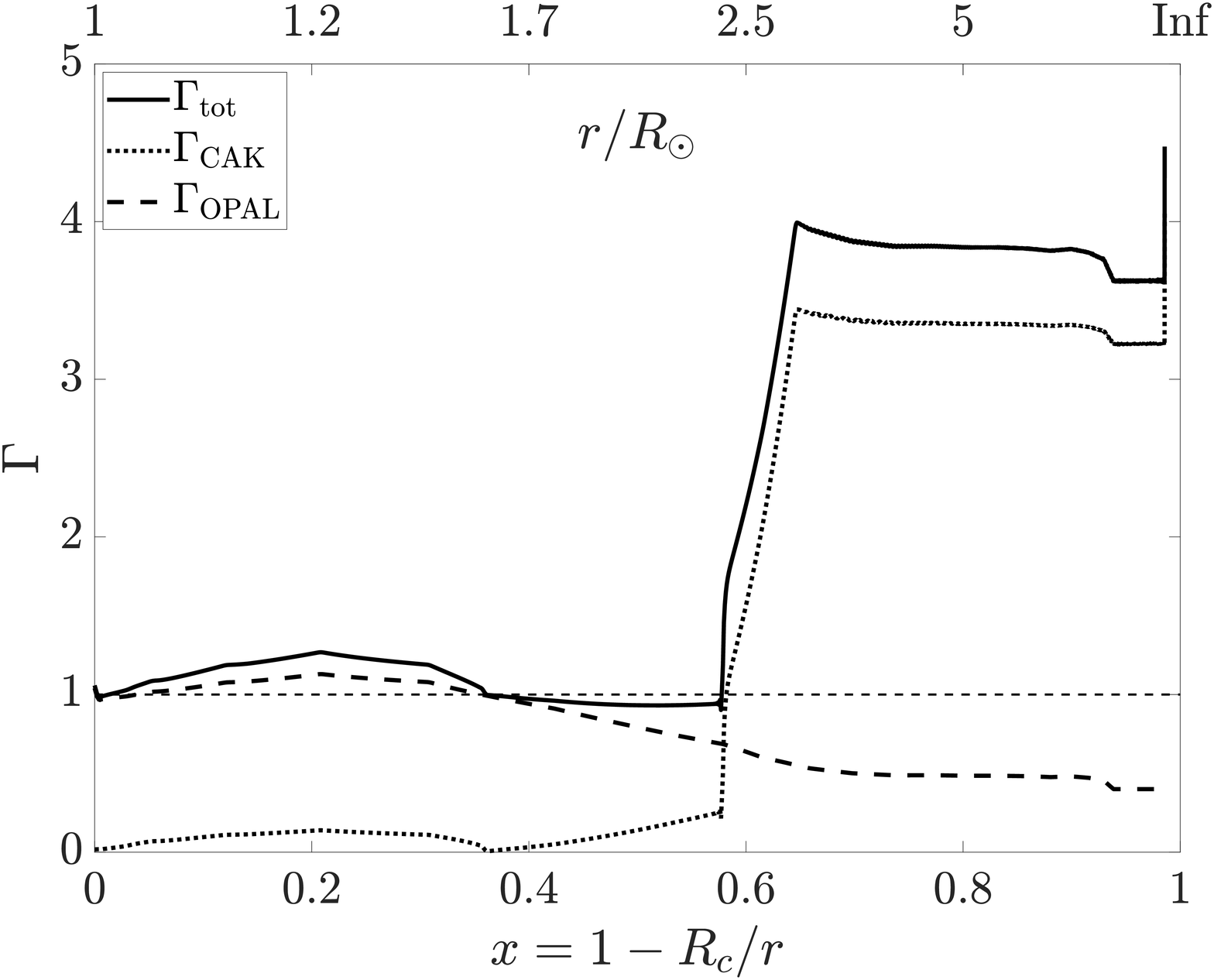}
    \caption{}
    \label{Fig_Time_dependent_gam_final}
  \end{subfigure}%
  \hfill%
  \begin{subfigure}[t]{.48\textwidth}
    \centering
    \includegraphics[width=\textwidth]{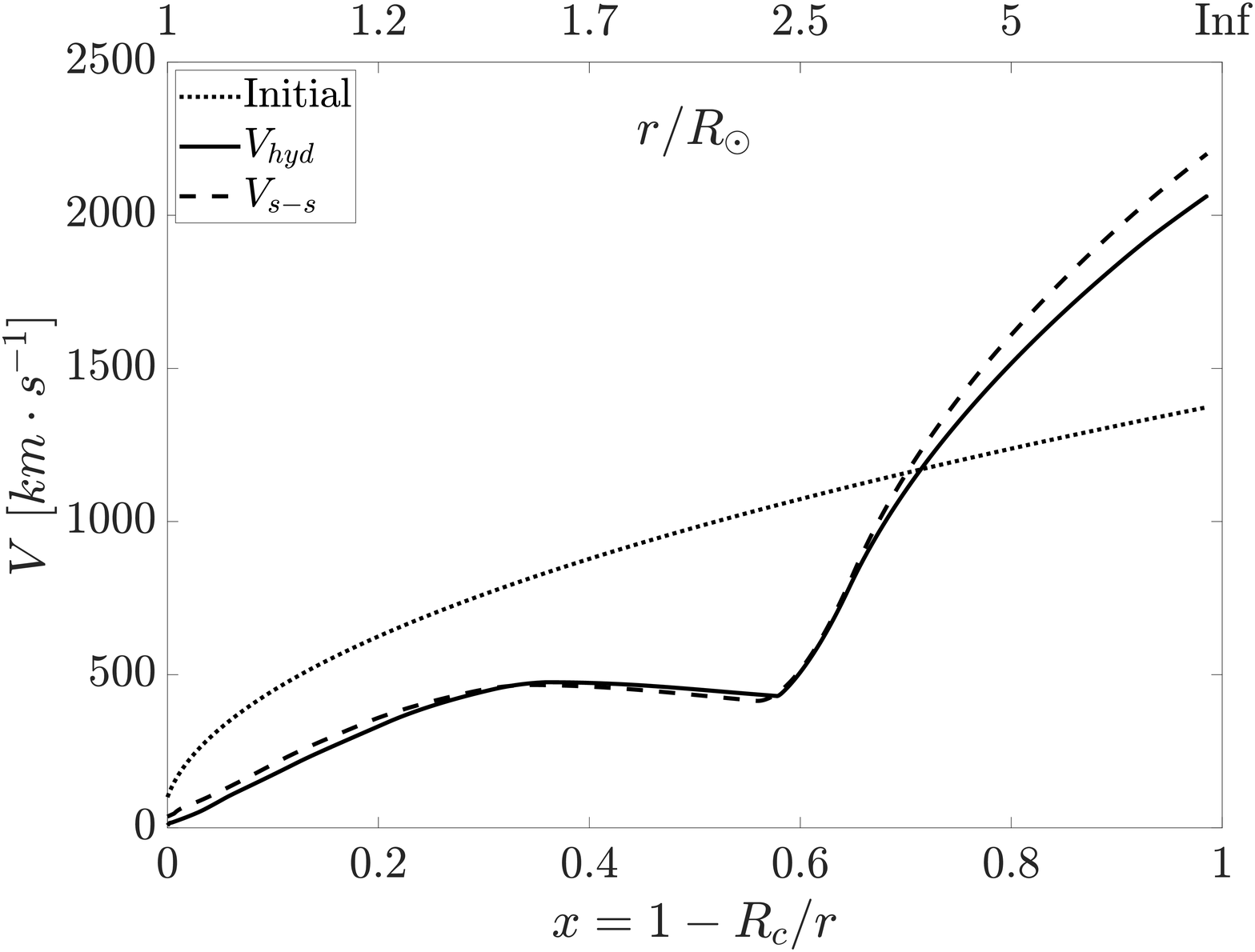}
    \caption{}
    \label{Fig_Time_dependent_velo_final}
  \end{subfigure}%
  \caption{Profiles of final time-step from the time-dependent numerical model (see Section \ref{Se_Time_dependant_model}). \textit{(a)} Eddington ratio, with the definition of line-styles the same as on Fig. \ref{Fig_steady_gam}. \textit{(b)} Comparison of the initial (dotted line), final time-step (solid line) and steady-state velocity fields.}
\end{figure*}

Running the simulation for in total $600\Unittime$, figures \ref{Fig_Time_dependent_gam_final} and \ref{Fig_Time_dependent_velo_final} show the Eddington ratio $\Gamma$ and the velocity profile of the final time-step.
From \newtxt{Figure} \oldtxt{figure}  \ref{Fig_Time_dependent_gam_final} we observe that, just as in the steady-state model, the \CAK-like force near the lower boundary on average only has a small contribution to the total force; contrarily, near the outer boundary, it is dominant. The radiation force from the \opal contribution has the opposite behaviour, dominating at the sonic\newtxt{-}point and in the inner regions, but becoming secondary in the outer parts. This then again leads to a velocity profile (solid line in Fig. \ref{Fig_Time_dependent_velo_final}) with a non-monotonic behaviour, which clearly can not be fit with the simple $\beta$-type velocity law assumed in the initial conditions (dotted line). 

This velocity profile can be directly compared to that found in our corresponding steady-state model, illustrated by the dashed line in Fig. \ref{Fig_Time_dependent_velo_final}\newtxt{, which shows} \oldtxt{and showing} close agreement between the two. The mass loss of the time-dependent model is $\dot M = 1.64\cdot10^{-5}\UnitMd$, which also agrees well with the mass-loss rate of $\dot M = 1.47\cdot10^{-5}\UnitMd$ from our steady-state model (see Section \ref{Se_steady_state}). In this way, our time-dependent models demonstrate that the 
'dynamically inflated' solution discussed in the previous section indeed works as a stable attractor for the stellar wind envelope, confirming our basic conclusions regarding a wind outflow controlled primarily by \opal opacities in the inner regions, and by the \CAK-like force in the outer. The small differences in velocity magnitude and mass loss between steady-state and time\newtxt{-}dependent models are mainly related to differences in setting the lower boundary. In steady-state computations we define the sonic-point to be exactly at the lower boundary, whereas in time-dependent computations it is allowed to self-adjust so that the sub-sonic part is also resolved. 

\subsection{$\dot M$ dependency on \cak \newtxt{$\alpha$} \oldtxt{force parameters}}\label{Se_Time_dependant_lineforce_dependancy}

A general finding of our baseline simulations in \newtxt{Sections} \oldtxt{Sects.} \ref{Se_steady_state} and \ref{Se_Time_dependant} is that (as long as the wind can escape), the mass-loss rate is primarily controlled by the conditions in the lower wind; in contrast, the terminal wind speed is controlled mainly by the outer wind conditions. Moreover, while it has turned out that a steady\newtxt{-state} model, such as the one presented in Section \ref{Se_steady_state}, is quite difficult to converge, the time-dependent simulations of this section are computationally somewhat easier to handle. 

Therefore, \oldtxt{by simply varying $\alpha_{\min}$ and $\alpha_{\max}$} we have used our time-dependent set-up to further explore the impact of the \CAK-like force parameters on the predictions of $\dot{M}$ and $v_\infty$. \newtxt{We investigate the impact of these parameters by varying only  $\alpha_{\min}$ and $\alpha_{\max}$, however we  point out that 
comparable effects could also be achieved by instead introducing corresponding radial variations of the second line-force parameter $\bar Q$.} To this end, we test a range of values $\alpha_{\max}\in[0.60,\, 0.66]$ and $\alpha_{\min}\in[0.48,\, 0.53]$, corresponding roughly to the scatter of $\alpha$ observed in the MC simulations displayed in Fig. \ref{Fig_steady_Springman_alpha}. The results of these tests are summarised in Tables \ref{Tab_alpha_min} and \ref{Tab_alpha_max}, along with the mass-loss rates and terminal velocities found for the different model-runs. In addition, we also tested variations in the spatial position of the $\alpha_{\max}$ to $\alpha_{\min}$ transition for the standard \cak force parameters in the range $x\in [0.3,\, 0.7]$. However, for these tests, no significant influence on the mass-loss rates of the models have been observed. Therefore, the following subsections focus exclusively on varying the values of $\alpha_{\max}$ and $\alpha_{\min}$. 

\subsubsection{Dependency on $\alpha_{\min}$}\label{Se_Time_dependant_lineforce_dependancy_amin}
As $\Gamma_\cak $ is the prevalent driver in the outer wind, one might initially expect a similar dependence of the mass-loss rate on $\alpha_{\min}$ as given by the standard \cak  model.
The \cak  expression for the mass-loss rate $\dot M_\cak $, typically derived from critical-point analyses for $\Gamma_\tot = \Gamma_e + \Gamma_\cak $ in the radial streaming limit, is:
\begin{equation}\label{Eq_steady_state_Mcak}
\dot M_\cak  = \frac{\L}{c^2}\frac{\alpha}{1-\alpha}\left(\frac{\bar Q \Gamma_e}{1-\Gamma_e}\right)^{\frac{1-\alpha}{\alpha}}\;.
\end{equation}
From this expression, it directly follows that $\dot{M}_\cak $ increases steeply if $\alpha$ is decreased while keeping all other parameters constant. However, it can be seen from Table~\ref{Tab_alpha_min} that in our models $\dot M$ varies only marginally across the range of considered $\alpha_{\min}$ values. This distinguishes these optically thick WR simulations from standard \CAK-type models applied to the optically thin winds of O-type stars. 
We further confirm the finding from the steady models that if $\alpha_{\min}$ is increased too much, the line force becomes insufficient to sustain the outflow. This happens because the line acceleration in the outer wind must be sufficient to prevent fall-back of the mass-loss initiated at the Fe-opacity bump. That is, the line acceleration must provide enough additional energy such that the previously introduced $\mathcal{W} \geq 0$ condition is fulfilled. As alluded to above, the \cak mass-loss rate formula cannot be directly used to predict the mass loss of this hybrid model. Nevertheless, the limiting value of $\alpha_{\min}$ that can sustain the wind can be estimated by inverting the above equation, solving for the $\alpha$ at which the \cak  mass-loss rate equals that of our actual model. For the given model with $\dot M = 1.64\cdot 10^{-5}\UnitMd$, one finds $\alpha_{\min} \approx 0.52$. The results of our numerical simulations, summarised in Table \ref{Tab_alpha_min}, indeed suggest that this provides a reasonable estimate for the limiting $\alpha_{\min}$ value.

This is also supported by closer inspection of the simulations with $\alpha_{\min} = 0.52$, which results in a flattening of the wind velocity profile as $\mathcal{W} \nless 0$. With a further increase to $\alpha_{\min} = 0.53$, we find that the \CAK-like force is unable to keep up with the mass flux initiated at the lower boundary. 
In the 1D simulations presented here, this then leads to a similar "failed wind" situation as discussed in the previous section; we discuss potential implications of this behaviour for multi-dimensional versions of our simulations at the end of the next section.

Finally, as summarised in Table \ref{Tab_alpha_min} the terminal velocity indeed is very sensitive to small changes of $\alpha_{\min}$. Overall, this thus suggests that within this formalism, $\dot{M}$ and $v_\infty$ are determined quite independently. 
As discussed further in Section \ref{Se_discussion_emperical}, such a basic decoupling of $\dot{M}$ and $v_\infty$ indeed seems to be supported by observational data of WR winds \citep[e.g. see][]{Hamann_19,Sander_19}. On the other hand, this result also seems to contradict that of the most recent study from \citet{Sander_20b}, who find a correlation between $\dot{M}$ and $v_\infty$. We further discuss possible reasons for this in Section \ref{Se_discussion_connection_Sander}. 

\begin{table}[ht]
\centering
\caption{Summary of mass-loss rates and terminal wind speeds for models with different values of $\alpha_{\min}$. The columns provide $\alpha_{\min}$ (value of $\alpha$ parameter at the outer boundary), $\alpha_{\max}$ (value of $\alpha$ parameter at the inner boundary), the mass-loss rate of the model $\dot M/10^{-5}$ in $\UnitMd$, the \cak mass loss $\dot M_\cak /10^{-5}$ in $\UnitMd$ corresponding to $\alpha_{\min}$, and the terminal velocity $v_\infty$ in $\UnitV$. The model with $\alpha_{\min} = 0.53$ failed to provide sufficient force to drive the outflow to the outer boundary, leading to termination of the numerical simulation (see text). }
\begin{tabular}{c c c c c}
\hline
\multirow{2}{*}{$\alpha_{\min}$ } & \multirow{2}{*}{$\alpha_{\max}$ } & $\dot M/10^{-5}$ & $\dot M_\cak /10^{-5}$ & $v_\infty$ \\
                                 &                                  & \multicolumn{2}{c}{$\UnitMd$} & $\UnitV$ \\
\hline

0.48 & \multirow{4}{*}{0.66} & 1.59 & 4.60 & 2765 \\ 
0.50 &                       & 1.64 & 2.70 & 2063 \\
0.52 &                       & 1.68 & 1.66 & 1279 \\
0.53 &                       & ---  & 1.33 & ---  \\
\hline
\end{tabular}
\label{Tab_alpha_min}
\end{table}

\subsubsection{Dependency on $\alpha_{\max}$}\label{Se_Time_dependant_lineforce_dependancy_amax}

In contrast to $\alpha_{\min}$, $\alpha_{\max}$ has little to no effect on determining the terminal velocity of the outflow. Nevertheless, it can affect $\dot{M}$. This is illustrated in Table \ref{Tab_alpha_max}, where a modest increase in $\dot{M}$ is found for decreasing values of $\alpha_{\max}$. We note, however, that again the magnitude of the mass loss variation is small in comparison to that seen in \CAK-type wind models of optically thin O-stars. 

This rather weak dependence on the \cak force stems from the fact that the dominant contribution to the total radiation force comes from the \opal opacities in the regions near the stellar core (see also Section \ref{Se_steady_state}). At the sonic point for $\alpha_{\max} = 0.6$, we find  $\Gamma_\cak  \approx 0.14$ and for the $\alpha_{\max} = 0.66$ we find $\Gamma = 0.02$, which is much lower even than our initial order of magnitude estimate $\Gamma \approx 0.1$ in Section \ref{Se_steady_state_missing_force}. This further supports that the \cak force has a negligible contribution at the sonic point. Moreover, since those inner parts accumulate the majority of the integrated wind optical depth, they also primarily control the lower-boundary temperature to which the overlying wind adjusts, and thereby also the mass-loss rate.   

\begin{table}[ht]
\centering
\caption{Summary of mass-loss rates and terminal wind speeds for different values of $\alpha_{\max}$. The columns are as in Table \ref{Tab_alpha_min}.}
\begin{tabular}{c c  c c}
\hline
\multirow{2}{*}{$\alpha_{\min}$ } & \multirow{2}{*}{$\alpha_{\max}$ } & $\dot M/10^{-5} $  & $v_\infty$ \\
     &    & $\UnitMd$ & $\UnitV$ \\
\hline
\multirow{3}{*}{0.50} 
  &  0.60  & 2.13 & 1787 \\
  &  0.63  & 1.84 & 1937 \\
  &  0.66  & 1.64 & 2063 \\
\hline
\end{tabular}
\label{Tab_alpha_max}
\end{table}

\section{Discussion}\label{Se_discussion}

\subsection{Implication for core radius problem}\label{Se_discussion_core_radius}

The dynamic models presented in \newtxt{Sections} \oldtxt{Sects.} \ref{Se_steady_state} and \ref{Se_Time_dependant} provide a natural explanation to the so-called "core radius problem" of classical WR stars (see Section \ref{Se_introduction}). To understand this, we introduce the standard Rosseland mean optical depth scale 
\begin{equation}\label{Eq_tau_ph}
    \tau_{\rm Ros} = \int\limits^{\infty}_{r} \kappa_{\rm Ros}\,\rho\,\d r = \int\limits^{\infty}_{r} \kappa_\opal\,\rho\,\d r\,,
\end{equation}
where $\kappa_{\rm Ros}$ is computed using the \newtxt{Rosseland mean opacities from the} \opal tabulation. 

\begin{figure}[ht]
\centering
\includegraphics[scale=0.19]{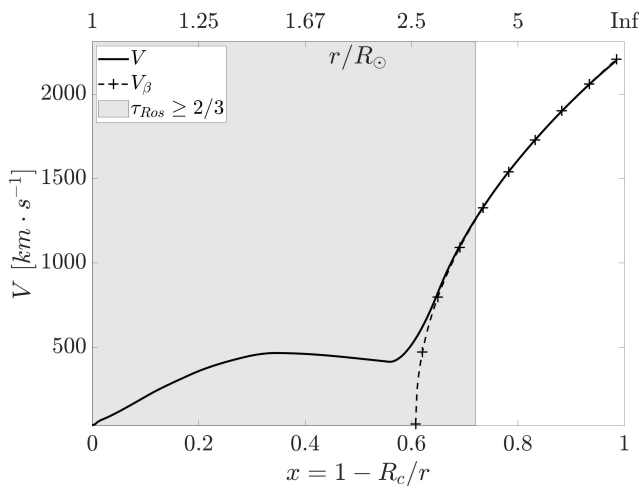}
\caption{The figure demonstrates a steady-state velocity profile (in solid-black) of the stellar model used in this paper. The grey shading marks the region of high photospheric continuum optical depth $\tau_{\rm Ros} \geq 2/3$, with $R(\tau_{\rm Ros} = 2/3) \approx 3.5R_\star$. In crossed-dashed line a $v_\beta$  velocity fit to the optically thin region is given, which was extrapolated to the optically thick region to recover the stellar core radius.}
\label{Fig_numerical_snap_fit}
\end{figure}

Then, following the standard spectroscopic model-approach of fitting an assumed $v_\beta$ velocity profile to the entire supersonic region, Fig. \ref{Fig_numerical_snap_fit} directly illustrates how the hydrostatic core $R_{\rm c}^{\rm sp}$ (where $v_\beta$ reaches sub-sonic velocities) in that case would be located at a radius $R_{\rm c}^{\rm sp} > R_{\rm c}$. Indeed, the core radius recovered in this way is a factor $\approx 2.6$ times larger than the true value, demonstrating the basic challenge of  determining WR hydrostatic radii from spectroscopic data \citep[e.g.,][]{Crowther_07}. The discrepancy essentially stems from the fact that $\tau_{\rm Ros} = 2/3$ is reached already at very high radii (see the grey area in Fig. \ref{Fig_numerical_snap_fit}), so that only the outermost velocity law is visible to us. Observational estimates of the hydrostatic stellar core radius must thus be obtained by some extrapolation of the velocity law assumed in the spectroscopic model, without any knowledge of the "true" velocity field in the invisible inner regions.

The dynamic models presented in this paper  explain this long-standing discrepancy simply by their extended ("dynamically inflated") region of low-velocity outflowing material above the hydrostatic stellar core.  

\begin{figure}[ht]
\centering
\includegraphics[scale=0.19]{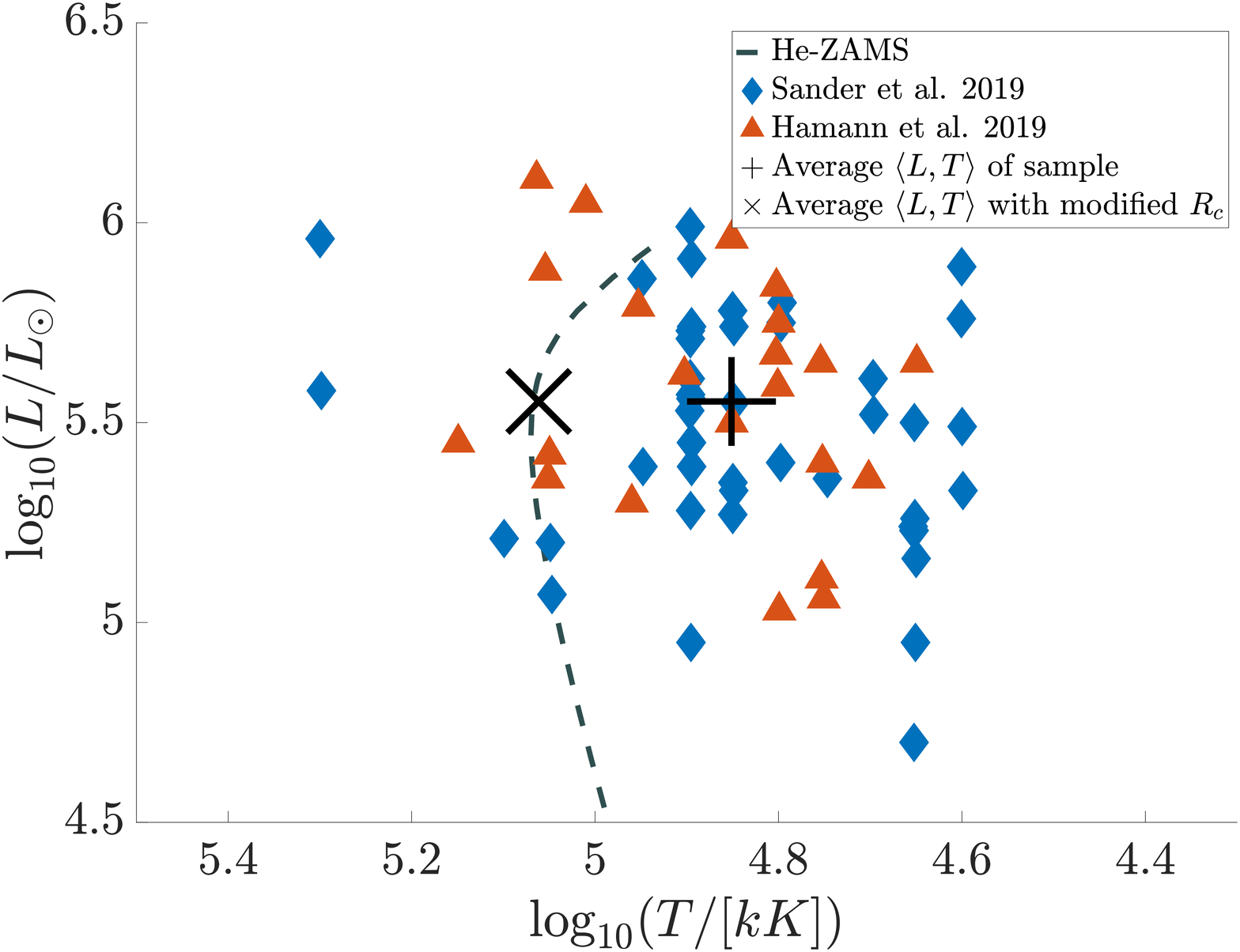}
\caption{Hertzsprung–Russell diagram including galactic WR stars of nitrogen (WN), carbon (WC), and oxygen sequences (WO). The solid shapes mark inferred core effective temperatures and luminosities from \citet{Hamann_19} (WN) and \citet{Sander_19}(WO,WC); the mean value is marked with a {\sffamily +}. The {\sffamily x} marks the location of the modified average luminosity-temperature, where the spectroscopically inferred core radius has been reduced
by a factor of 3 (see text). The grey dashed line shows the approximate predicted location of the Helium Zero Age Main Sequence (He-ZAMS) computed using MESA (see the text).}
\label{Fig_HRD_population}
\end{figure}

To further illustrate this effect, let us inspect the observational positioning of classical WR stars on the Hertzsprung–Russell diagram (HRD), where the stars are placed according to their observationally inferred luminosities and hydrostatic core effective temperatures.  Figure  \ref{Fig_HRD_population} here shows the HRD where the observed galactic sample of WR stars by \citet{Hamann_19} and \cite{Sander_19} is placed. Positions of these stars on the HRD can then be compared to the location of the so-called Helium Zero Age Main Sequence (He-ZAMS) (dashed line in figure), computed here with MESA using the standard mixing-length theory for convection with solar metallicity and relative metal fractions as given by \citet{Asplund_09}\footnote{The computed location of the He-ZAMS as well as Input files to reproduce our MESA simulations are provided at \url{https://doi.org/10.5281/zenodo.4054811}}. On average, observationally inferred positions of WR stars are then expected to lay to the left of this He-ZAMS line. This is because typical stellar evolution models, where an additional energy transport is typically also invoked in order to prevent the WR cores from inflating (see Section \ref{Se_steady_state_dynamic_inflation}), occupy the region to the hotter side of He-ZAMS on the HRD. Figure \ref{Fig_HRD_population}, however, shows that the average position of these WR stars (indicated by \texttt{+} in the figure) is located on the cooler side of the He-ZAMS instead. To improve the agreement between spectroscopy and evolution predictions a simple correction factor $\approx 3$ can be applied to the spectroscopic radii, which then shifts the 
core effective temperature location (now identified on figure by \texttt{x}). And indeed, such a correction factor matches quite well that implied by our "dynamic inflation" models above.

\subsection{Empirical constraints and model dependencies  of mass loss and terminal velocity}\label{Se_discussion_emperical}

\begin{figure}[ht]
\centering
\includegraphics[scale=0.19]{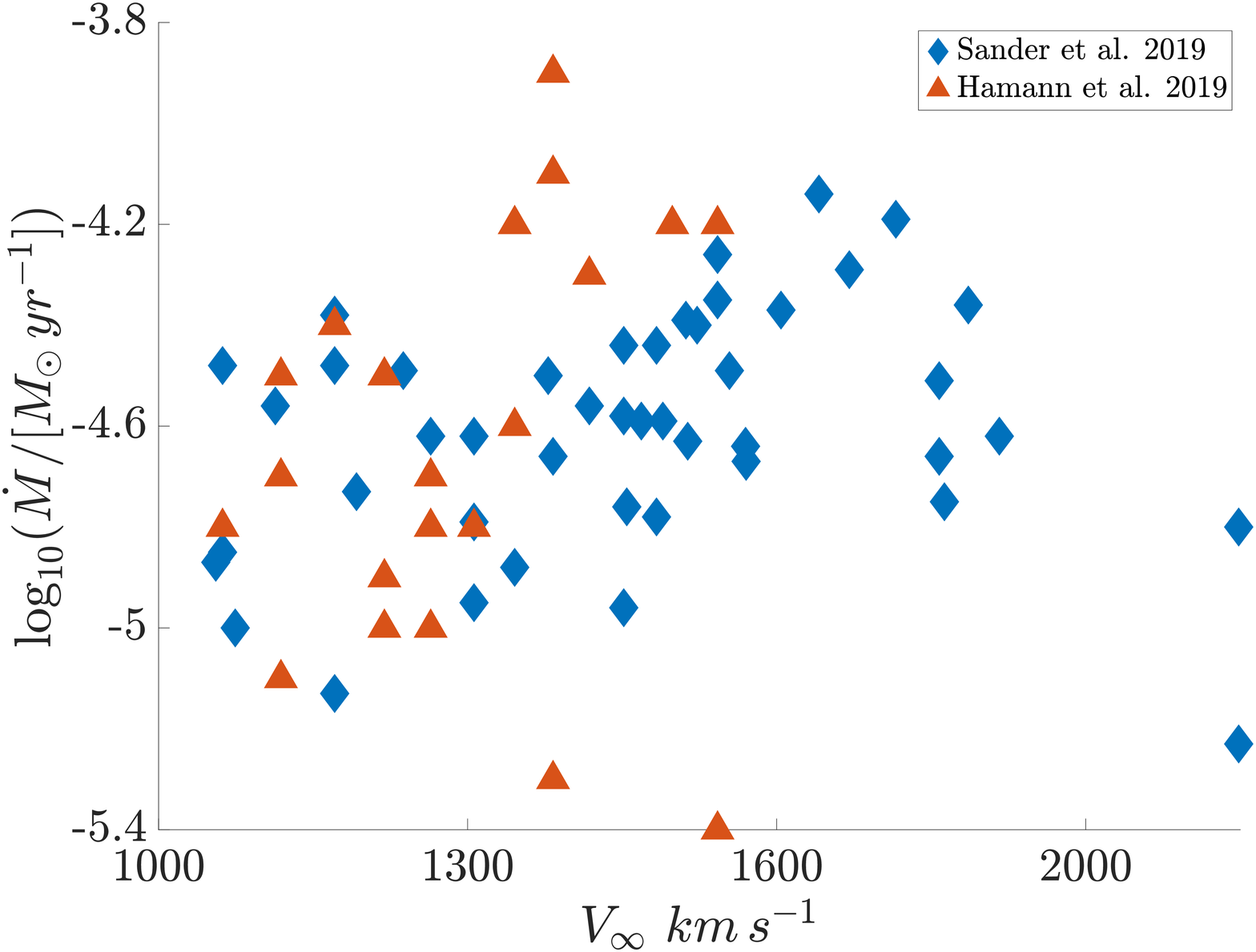}
\caption{The distribution of mass-loss rates and terminal velocities derived from the spectroscopic fitting by \citet{Sander_19} and \citet{Hamann_19} for an observed stellar population of galactic WR stars of nitrogen (WN), carbon (WC), and oxygen sequences (WO). Mass-loss rates and terminal velocities derived in our steady-state  ($\log_{10}(\dot M/\UnitMd) = -4.83$, $v_\infty = 2200 \UnitV$)  and time-dependent ($\log_{10}(\dot M/\UnitMd) = -4.78$, $v_\infty = 2063\UnitV$) simulations are consistent  with the range of mass loss shown on this figure.} 
\label{Fig_Md_V_fit}
\end{figure}

The basic model presented in the previous sections shows an overall quite good agreement with empirical constraints on terminal velocities and mass-loss rates of WR stars. Figure \ref{Fig_Md_V_fit} plots such empirically inferred values of $\dot M$  and $v_\infty$, again from \citet{Hamann_19} and \citet{Sander_19}. Compared to the mass-loss rates and terminal velocities predicted by our hydrodynamic and steady-state models (summarised in Table \ref{Tab_model_params}), the latter is located at the high end of the inferred values. However, since $v_\infty$ is primarily determined by the line-force in the outer wind regions, which here is parametrised by the simple radial streaming \CAK-approximation outlined in previous sections, a more detailed approach will be necessary to make more quantitative comparisons to empirical $v_\infty$. 
Inspection of Fig. \ref{Fig_Md_V_fit} also reveals that the empirical mass-loss rates and terminal wind speeds show no prominent anti-correlation.
This further supports another basic characteristic of our models, namely that $\dot M$ and $v_\infty$ are quite decoupled (see Section \ref{Se_Time_dependant_lineforce_dependancy_amin}).

\begin{table}[ht]
    \centering
    \caption{Summary of derived parameters of our steady-state and time dependent models. Listed are mass-loss rater, terminal velocity, radius of photosphere, and effective temperature at photosphere. }
    \begin{tabular}{ c c c c c }
         \hline
         \multirow{2}{*}{Model} & $\dot M$ & $v_\infty$ & $R_{\rm ph}$ & $T_{\rm ph,\, eff}$\\
            & $\rm M_\odot\, yr^{-1}$ & $\rm km\,s^{-1}$ & $R_\odot$ & $\rm kK$\\
         \hline
         Steady-state & $1.47\cdot 10^{-5}$ & $2200$ & $4.48$ & $62$ \\
         Time dependent & $1.64\cdot 10^{-5}$ & $2063$ & $4.64$ & $60$ \\
         \hline
    \end{tabular}
    \label{Tab_model_params}
\end{table}

\subsection{Dependence on metallicity}
\label{Se_discussion_metallicity}
The model of dynamic inflation presented in this paper relies on the fact that the $\Gamma = 1$ condition is reached already at the Fe-opacity bump. As such it is sensitive to the variation of opacity at the Fe-opacity bump, which is directly controlled by metallicity \citep[e.g.,][]{Cantiello_09}. Qualitatively, the model of dynamic inflation should respond to a decrease/increase in metallicity by reducing/enlarging the scale of inflation and the associated mass loss. Moreover, at low enough metallicity the $\Gamma = 1$ condition will not be fulfilled at the Fe-bump, in which case dynamic inflation will no longer take place from the hot Fe-bump, and a standard optically thin wind \newtxt{develops} \oldtxt{develop} instead. This behaviour is also in line with the recently published models by \citet{Sander_20b}.

\subsection{Comparison to co-moving frame models} 
\label{Se_discussion_connection_Sander}

As mentioned above, recently \citet{Sander_20a,Sander_20b} published \newtxt{time-independent} wind simulations of classical WR stars. These models solve the same steady-state e.o.m. as here\footnote{except that they also include a constant "turbulent velocity" part in their calculation of an "effective" sound speed}, however using a different method to calculate the radiative acceleration. There $\Gamma(r)$ is derived from full solutions of the frequency-dependent radiative transfer equation in the co-moving frame (CMF), without the use of any parameterisation to estimate the line-force contribution \citep[see also][for similar approaches applied to O-stars]{Sundqvist_19,Bjorklund_20}. Although these models thus compute $\Gamma(r)$ for a given velocity and density structure in a more detailed way than here, this CMF method (as currently implemented) has one major disadvantage. Because of the way the CMF transfer equation is numerically solved, it can only be applied to monotonically increasing (or decreasing) velocity fields $v(r)$. As such, this method cannot be used to derive $\Gamma(r)$ for the type of non-monotonic velocity structures found here (see Figs. \ref{Fig_steady_velo}, \ref{Fig_Time_dependent_velo_final}). 

The way \citet{Sander_20a} overcome this issue is by assuming an ad-hoc very high degree of "clumping" in the WR outflow \citep[see also][]{Grafener_05}. They parameterise this using a simple model where all wind mass is assumed to be concentrated within overdense clumps, such that the densities of these clumps are $\rho_{\rm cl} = \langle \rho \rangle / f_{\rm vol} = D \langle \rho \rangle$,
where $\langle \rho \rangle$ is the mean density (assumed to be preserved with respect to a smooth model), $f_{\rm vol}$ the volume fraction of the total wind contained by clumps, and $D$ the clump-overdensity factor.

By adopting a very high $D=50$ ($f_{\rm vol} =0.02$), the ionisation balance of the wind is shifted so that more effective driving lines become available, and the line-force is thereby increased enough to avoid any deceleration regions. When assuming instead a more modest $D=10$ in their models, they do experience the same type of $\Gamma <1$ regions as here, which leads then to a  non-monotonic $v(r)$ that cannot be handled by the CMF radiative transfer (see Fig. 1 in \citealt{Sander_20a}, and their corresponding discussion).
Since observations of electron scattering wings in WR outflows suggest quite modest values $D \approx 4-10$ \citep[see overviews by][]{Crowther_07,Puls_08}, and theoretical wind-clumping models following the line-driven instability \citep{Owocki_88} have so far only been developed for OB-stars \citep[with again typical values $D \approx 10$; ][]{Sundqvist_18,Driessen_19}, it is at present not clear how the high clump densities needed to keep $v(r)$ monotonic in CMF-based WR models might be physically justified. 

The models presented here are thus very complementary to such CMF-based simulations. \newtxt{Although our more simplified treatment of $\Gamma$ involves certain approximations (as discussed in previous sections) it allows us to model regions of stagnation and deceleration (see Sections \ref{Se_steady_state} and \ref{Se_Time_dependant}), whereas such non-monotonic flows are incompatible with the methods currently used in CMF-based calculations. Moreover, the method presented here allows for time-dependent simulations that otherwise would not be computationally feasible.} \oldtxt{Due to our more simplified treatment of $\Gamma$, we can readily handle non-monotonic velocity fields in both our steady and time-dependent simulations (see Sections \ref{Se_steady_state} and \ref{Se_Time_dependant}).} In this respect, we also recall our finding that if the line-force in the outer wind is decreased too much, the wind not only experiences regions of deceleration but may eventually reach zero velocity. In the 1D simulations presented here, this leads to "failed wind solutions" and model-termination. In a multi-dimensional simulation, one may speculate that it \oldtxt{rather} might \newtxt{instead} result in a very complex wind structure of co-existing regions of upflows and inflows. In turn, this may then cause shocks and perhaps increased levels of clump-formation. \newtxt{These could be simulated in multi-dimensional hydrodynamics simulations using the method we present here.}

The models presented here also demonstrate that even with multiple extended regions where $\Gamma < 1$, the accumulated net energy of the wind can still be positive (i.e., we can still have  $\mathcal{W}\geq 0$). This suggests that the WR outflow can be driven without invoking such a high degree of clumping, but instead allowing for non-monotonic velocities. For example, let us consider the lower panel of \newtxt{Fig.}\oldtxt{figure} 1 in \citet{Sander_20a}, where the authors provide $\Gamma(r)$ not only for their final $D=50$ model (their solid line) but also for their collapsed $D=10$ solution (their dashed line). From these profiles, we estimated the net energy corresponding to both models and found that, indeed, both of the force configurations provide $\mathcal{W} > 0$. As such, both solutions may actually escape the stellar potential, but since the $D=10$ model would have contained regions of negative acceleration it could not be treated by the assumed CMF radiative transfer framework.

Note that \citet{Sander_20a} make the argument that although the assumed very high degree of clumping influences $v_\infty$ significantly, it does not have a major impact on $\dot{M}$, since clumping is assumed to be present only at supersonic velocities. This is quite consistent with our finding here. On the other hand, \citet{Sander_20b} also find a correlation between $\dot{M}$ and $v_\infty$, which we do not necessarily observe in our models. This result, however, might simply be related to them forcing a monotonic velocity field in their simulations, since the same correlation was also found in the $\beta$-law WR models by \citet{Grafener_17}.

\section{Summary and Future Work}\label{Se_summary_future}

This paper has presented a model for the wind outflows of classical Wolf-Rayet stars, based on a hybrid opacity method where \opal opacities are used in combination with a \CAK-like line force parametrisation. The latter is used here to account for the effects of Doppler shifts upon spectral line opacity in an approximate way. Our hybrid opacity model essentially results in a "two stage" WR stellar wind: a supersonic outflow is first initiated in optically thick layers (at the so-called "Fe-bump") by the \opal opacities. However, these opacities are not able to prevent fallback upon the core, and so it is the \CAK-like force that takes over the driving in the outer regions and ensures that the initiated outflow can also escape the stellar potential. 

Our approach leads to deep-seated wind initiation with high mass-loss rates and terminal velocities (see Table \ref{Tab_model_params}), in the range of values typically inferred for classical WR stars by spectroscopic analyses. The simulations display dense and slow -- "dynamically inflated" --  sub-photospheric layers, offering a natural explanation to the so-called "core-radius" problem of classical WR stars. 
Direct comparison with hydrostatic computations further shows that neglecting the dynamical terms in the e.o.m. instead leads to drastic inflation of the then assumed static envelope. In this way, we demonstrate explicitly that the need to invoke huge amounts of ad-hoc energy transport in these convectively inefficient regions of stellar structure and evolution calculations is simply an artefact of neglecting the dynamic outer boundary. 
 
The paper here uses a line-force parameterisation that introduces a very simple radial variation to the \cak power-law index $\alpha$ in order to sustain the outer wind outflow. The quite simple nature of this basic set-up means that it should be rather straightforward to extend these first simulations toward exploration of a more complete stellar parameter space for classical WR stars. Since the assumed \cak  line-force parameterisation proved to have little effect on the resulting mass loss (as long as we ensure a positive wind energy), such model-grids might then offer good alternatives for direct inclusion of mass-loss rates into stellar evolution models. 

The assumed treatment of $\alpha$ does influence the outer wind velocity profile and terminal speed. As such, constraining the morphology of the re-accelerating part of the outflow will require more detailed studies. This is a very challenging task, however, since a non-parametrised calculation of the line-force in these regions must be able to deal with non-monotonic velocity fields. As such, current CMF radiative transfer methods, like those applied in some alternative WR- \citep{Grafener_05, Sander_20a, Sander_20b} and O-type star wind models \citep{Sundqvist_19, Bjorklund_20}, will not be sufficient.   

The assumed variation of $\alpha$ can also be varied in multi-dimensional extensions of the time-dependent simulations presented here. In such a study we may allow parts of the wind material to fall back into the deep regions of the outflow.  While in 1D simulations such fallback eventually always leads to model termination, in a multi-dimensional setting it may instead create a highly structured, turbulent flow. This might then still deposit enough momentum into the plasma that a wind can be sustained, but now with clumps of material perhaps mechanically transporting some of the momentum through collisions. 

As the wind velocity field also affects spectral line formation, another important follow-up regards analysis of spectral features associated with the non-monotonic velocities found in this paper. In particular, a key question becomes if there are any fundamental differences in the predicted strengths and shapes of strategic spectral lines calculated from our models as compared to those predicted by standard spectroscopic models assuming a monotonic $\beta$-type velocity law \citep[e.g., \texttt{cmfgen},][]{Hillier_98}. Using the newly developed 3D radiative transfer code by \citet{Hennicker_20}, studies of this are already underway and will be presented in an upcoming paper.  
Finally, a very interesting problem to address concerns stellar structure and evolution, and the manifestation of dynamic inflation in different evolutionary stages. Such a study might be especially important since quite many stars (some already at earlier evolution stages) reach the Eddington limit at opacity bumps in regions that might have insufficient convective energy transport. As demonstrated in Section \ref{Se_steady_state}, dynamically inflated envelopes should then develop, which in turn could have a significant feedback on the stellar structure as well as on the further evolution of the star.

\begin{acknowledgements} 
The authors would like to thank J. Puls, T. Shenar and all members of the KUL EQUATION group for fruitful discussion, comments, and suggestions.  
LP, JS, NK, LD, AdK, LM, and HS acknowledge support from the KU Leuven C1 grant MAESTRO C16/17/007.  
LP further acknowledges the additional support from Shota Rustaveli Georgian National Foundation Grant Project No. FR17-391. JS further
acknowledges additional support by the Belgian Research Foundation Flanders (FWO)
Odysseus program under grant number G0H9218N.
\end{acknowledgements}

\bibliographystyle{aa} 
\bibliography{biblio} 

\begin{thebibliography}{62}
\expandafter\ifx\csname natexlab\endcsname\relax\def\natexlab#1{#1}\fi

\bibitem[{{Asplund} {et~al.}(2009){Asplund}, {Grevesse}, {Sauval}, \&
  {Scott}}]{Asplund_09}
{Asplund}, M., {Grevesse}, N., {Sauval}, A.~J., \& {Scott}, P. 2009, \araa, 47,
  481

\bibitem[{{Beals}(1929)}]{Beals_29}
{Beals}, C.~S. 1929, \mnras, 90, 202

\bibitem[{{Bj{\"o}rklund} {et~al.}(2020){Bj{\"o}rklund}, {Sundqvist}, {Puls},
  \& {Najarro}}]{Bjorklund_20}
{Bj{\"o}rklund}, R., {Sundqvist}, J.~O., {Puls}, J., \& {Najarro}, F. 2020,
  arXiv e-prints, arXiv:2008.06066

\bibitem[{{Cantiello} {et~al.}(2009){Cantiello}, {Langer}, {Brott}, {de Koter},
  {Shore}, {Vink}, {Voegler}, {Lennon}, \& {Yoon}}]{Cantiello_09}
{Cantiello}, M., {Langer}, N., {Brott}, I., {et~al.} 2009, \aap, 499, 279

\bibitem[{{Cassinelli}(1991)}]{Cassinelli_91}
{Cassinelli}, J.~P. 1991, in IAU Symposium, Vol. 143, Wolf-Rayet Stars and
  Interrelations with Other Massive Stars in Galaxies, ed. K.~A. {van der
  Hucht} \& B.~{Hidayat}, 289

\bibitem[{{Castor} {et~al.}(1975){Castor}, {Abbott}, \& {Klein}}]{Castor_75}
{Castor}, J.~I., {Abbott}, D.~C., \& {Klein}, R.~I. 1975, \apj, 195, 157

\bibitem[{Courant {et~al.}(1928)Courant, Friedrichs, \& Lewy}]{Courant_28}
Courant, R., Friedrichs, K., \& Lewy, H. 1928, Mathematische Annalen, 100, 32

\bibitem[{{Crowther}(2007)}]{Crowther_07}
{Crowther}, P.~A. 2007, \araa, 45, 177

\bibitem[{{Crowther} {et~al.}(2010){Crowther}, {Schnurr}, {Hirschi}, {Yusof},
  {Parker}, {Goodwin}, \& {Kassim}}]{Crowther_10}
{Crowther}, P.~A., {Schnurr}, O., {Hirschi}, R., {et~al.} 2010, \mnras, 408,
  731

\bibitem[{{de Koter} {et~al.}(1997){de Koter}, {Heap}, \&
  {Hubeny}}]{deKoter_97}
{de Koter}, A., {Heap}, S.~R., \& {Hubeny}, I. 1997, \apj, 477, 792

\bibitem[{{Doran} {et~al.}(2013){Doran}, {Crowther}, {de Koter}, {Evans},
  {McEvoy}, {Walborn}, {Bastian}, {Bestenlehner}, {Gr{\"a}fener}, {Herrero},
  {K{\"o}hler}, {Ma{\'\i}z Apell{\'a}niz}, {Najarro}, {Puls}, {Sana},
  {Schneider}, {Taylor}, {van Loon}, \& {Vink}}]{Doran_13}
{Doran}, E.~I., {Crowther}, P.~A., {de Koter}, A., {et~al.} 2013, \aap, 558,
  A134

\bibitem[{{Driessen} {et~al.}(2019){Driessen}, {Sundqvist}, \&
  {Kee}}]{Driessen_19}
{Driessen}, F.~A., {Sundqvist}, J.~O., \& {Kee}, N.~D. 2019, \aap, 631, A172

\bibitem[{{Ekstr{\"o}m} {et~al.}(2012){Ekstr{\"o}m}, {Georgy}, {Eggenberger},
  {Meynet}, {Mowlavi}, {Wyttenbach}, {Granada}, {Decressin}, {Hirschi},
  {Frischknecht}, {Charbonnel}, \& {Maeder}}]{Ekstrom_12}
{Ekstr{\"o}m}, S., {Georgy}, C., {Eggenberger}, P., {et~al.} 2012, \aap, 537,
  A146

\bibitem[{{Gayley}(1995)}]{Gayley_95}
{Gayley}, K.~G. 1995, \apj, 454, 410

\bibitem[{{Gayley} {et~al.}(1995){Gayley}, {Owocki}, \&
  {Cranmer}}]{Gayley_Owocki_95}
{Gayley}, K.~G., {Owocki}, S.~P., \& {Cranmer}, S.~R. 1995, \apj, 442, 296

\bibitem[{{Gr{\"a}fener} \& {Hamann}(2005)}]{Grafener_05}
{Gr{\"a}fener}, G. \& {Hamann}, W.~R. 2005, \aap, 432, 633

\bibitem[{{Gr{\"a}fener} {et~al.}(2017){Gr{\"a}fener}, {Owocki}, {Grassitelli},
  \& {Langer}}]{Grafener_17}
{Gr{\"a}fener}, G., {Owocki}, S.~P., {Grassitelli}, L., \& {Langer}, N. 2017,
  \aap, 608, A34

\bibitem[{{Gr{\"a}fener} {et~al.}(2012){Gr{\"a}fener}, {Owocki}, \&
  {Vink}}]{Grafener_12}
{Gr{\"a}fener}, G., {Owocki}, S.~P., \& {Vink}, J.~S. 2012, \aap, 538, A40

\bibitem[{{Grassitelli} {et~al.}(2018){Grassitelli}, {Langer}, {Grin},
  {Mackey}, {Bestenlehner}, \& {Gr{\"a}fener}}]{Grassitelli_18}
{Grassitelli}, L., {Langer}, N., {Grin}, N.~J., {et~al.} 2018, \aap, 614, A86

\bibitem[{{Grevesse} \& {Noels}(1993)}]{Grevesse_93}
{Grevesse}, N. \& {Noels}, A. 1993, Physica Scripta Volume T, 47, 133

\bibitem[{{Groh} {et~al.}(2013){Groh}, {Georgy}, \& {Ekstr{\"o}m}}]{Groh_13b}
{Groh}, J.~H., {Georgy}, C., \& {Ekstr{\"o}m}, S. 2013, \aap, 558, L1

\bibitem[{{Hamann} {et~al.}(2019){Hamann}, {Gr{\"a}fener}, {Liermann},
  {Hainich}, {Sander}, {Shenar}, {Ramachand ran}, {Todt}, \&
  {Oskinova}}]{Hamann_19}
{Hamann}, W.~R., {Gr{\"a}fener}, G., {Liermann}, A., {et~al.} 2019, \aap, 625,
  A57

\bibitem[{Harten(1983)}]{Harten_83}
Harten, A. 1983, Journal of Computational Physics, 49,, 357

\bibitem[{{Hennicker} {et~al.}(2020){Hennicker}, {Puls}, {Kee}, \&
  {Sundqvist}}]{Hennicker_20}
{Hennicker}, L., {Puls}, J., {Kee}, N.~D., \& {Sundqvist}, J.~O. 2020, \aap,
  633, A16

\bibitem[{{Hillier} \& {Miller}(1998)}]{Hillier_98}
{Hillier}, D.~J. \& {Miller}, D.~L. 1998, \apj, 496, 407

\bibitem[{{Iglesias} \& {Rogers}(1996)}]{Iglesias_96}
{Iglesias}, C.~A. \& {Rogers}, F.~J. 1996, \apj, 464, 943

\bibitem[{{Ishii} {et~al.}(1999){Ishii}, {Ueno}, \& {Kato}}]{Ishii_99}
{Ishii}, M., {Ueno}, M., \& {Kato}, M. 1999, \pasj, 51, 417

\bibitem[{{Kee} {et~al.}(2016){Kee}, {Owocki}, \& {Sundqvist}}]{Kee_16}
{Kee}, N.~D., {Owocki}, S., \& {Sundqvist}, J.~O. 2016, \mnras, 458, 2323

\bibitem[{{Lamers} \& {Leitherer}(1993)}]{Lamers_93}
{Lamers}, H. J.~G.~L.~M. \& {Leitherer}, C. 1993, \apj, 412, 771

\bibitem[{{Lucy}(1971)}]{Lucy_71}
{Lucy}, L.~B. 1971, \apj, 163, 95

\bibitem[{{Lucy} \& {Abbott}(1993)}]{Lucy_93}
{Lucy}, L.~B. \& {Abbott}, D.~C. 1993, \apj, 405, 738

\bibitem[{{Lucy} \& {Solomon}(1970)}]{Lucy_70}
{Lucy}, L.~B. \& {Solomon}, P.~M. 1970, \apj, 159, 879

\bibitem[{{Nugis} \& {Lamers}(2002)}]{Nugis_02}
{Nugis}, T. \& {Lamers}, H.~J.~G.~L.~M. 2002, \aap, 389, 162

\bibitem[{{Owocki}(2014)}]{Owocki_14}
{Owocki}, S. 2014, arXiv e-prints, arXiv:1409.2084

\bibitem[{{Owocki} {et~al.}(1988){Owocki}, {Castor}, \& {Rybicki}}]{Owocki_88}
{Owocki}, S.~P., {Castor}, J.~I., \& {Rybicki}, G.~B. 1988, \apj, 335, 914

\bibitem[{{Owocki} {et~al.}(2017){Owocki}, {Townsend}, \&
  {Quataert}}]{Owocki_17}
{Owocki}, S.~P., {Townsend}, R. H.~D., \& {Quataert}, E. 2017, \mnras, 472,
  3749

\bibitem[{{Pauldrach} {et~al.}(1993){Pauldrach}, {Feldmeier}, {Puls}, \&
  {Kudritzki}}]{Pauldrach_93}
{Pauldrach}, A.~W.~A., {Feldmeier}, A., {Puls}, J., \& {Kudritzki}, R.~P. 1993,
  \ssr, 66, 105

\bibitem[{{Paxton} {et~al.}(2013){Paxton}, {Cantiello}, {Arras}, {Bildsten},
  {Brown}, {Dotter}, {Mankovich}, {Montgomery}, {Stello}, {Timmes}, \&
  {Townsend}}]{Paxton_13}
{Paxton}, B., {Cantiello}, M., {Arras}, P., {et~al.} 2013, \apjs, 208, 4

\bibitem[{{Paxton} {et~al.}(2019){Paxton}, {Smolec}, {Schwab}, {Gautschy},
  {Bildsten}, {Cantiello}, {Dotter}, {Farmer}, {Goldberg}, {Jermyn}, {Kanbur},
  {Marchant}, {Thoul}, {Townsend}, {Wolf}, {Zhang}, \& {Timmes}}]{Paxton_19}
{Paxton}, B., {Smolec}, R., {Schwab}, J., {et~al.} 2019, \apjs, 243, 10

\bibitem[{{Petrenz} \& {Puls}(2000)}]{Petrenz_00}
{Petrenz}, P. \& {Puls}, J. 2000, \aap, 358, 956

\bibitem[{{Petrovic} {et~al.}(2006){Petrovic}, {Pols}, \&
  {Langer}}]{Petrovic_06}
{Petrovic}, J., {Pols}, O., \& {Langer}, N. 2006, \aap, 450, 219

\bibitem[{{Prantzos} {et~al.}(2018){Prantzos}, {Abia}, {Limongi}, {Chieffi}, \&
  {Cristallo}}]{Prantzos_18}
{Prantzos}, N., {Abia}, C., {Limongi}, M., {Chieffi}, A., \& {Cristallo}, S.
  2018, \mnras, 476, 3432

\bibitem[{{Proga} {et~al.}(1998){Proga}, {Stone}, \& {Drew}}]{Proga_98}
{Proga}, D., {Stone}, J.~M., \& {Drew}, J.~E. 1998, \mnras, 295, 595

\bibitem[{{Puls} {et~al.}(2000){Puls}, {Springmann}, \& {Lennon}}]{Puls_00}
{Puls}, J., {Springmann}, U., \& {Lennon}, M. 2000, \aaps, 141, 23

\bibitem[{{Puls} {et~al.}(2005){Puls}, {Urbaneja}, {Venero}, {Repolust},
  {Springmann}, {Jokuthy}, \& {Mokiem}}]{Puls_05}
{Puls}, J., {Urbaneja}, M.~A., {Venero}, R., {et~al.} 2005, \aap, 435, 669

\bibitem[{{Puls} {et~al.}(2008){Puls}, {Vink}, \& {Najarro}}]{Puls_08}
{Puls}, J., {Vink}, J.~S., \& {Najarro}, F. 2008, \aapr, 16, 209

\bibitem[{{Ramachandran} {et~al.}(2018){Ramachandran}, {Hamann}, {Hainich},
  {Oskinova}, {Shenar}, {Sander}, {Todt}, \& {Gallagher}}]{Ramachandran_18}
{Ramachandran}, V., {Hamann}, W.~R., {Hainich}, R., {et~al.} 2018, \aap, 615,
  A40

\bibitem[{{Ro} \& {Matzner}(2016)}]{Ro_16}
{Ro}, S. \& {Matzner}, C.~D. 2016, \apj, 821, 109

\bibitem[{{Sander} {et~al.}(2019){Sander}, {Hamann}, {Todt}, {Hainich},
  {Shenar}, {Ramachandran}, \& {Oskinova}}]{Sander_19}
{Sander}, A.~A.~C., {Hamann}, W.~R., {Todt}, H., {et~al.} 2019, \aap, 621, A92

\bibitem[{{Sander} \& {Vink}(2020)}]{Sander_20b}
{Sander}, A. A.~C. \& {Vink}, J.~S. 2020, arXiv e-prints, arXiv:2009.01849

\bibitem[{{Sander} {et~al.}(2020){Sander}, {Vink}, \& {Hamann}}]{Sander_20a}
{Sander}, A. A.~C., {Vink}, J.~S., \& {Hamann}, W.~R. 2020, \mnras, 491, 4406

\bibitem[{{Sanyal} {et~al.}(2015){Sanyal}, {Grassitelli}, {Langer}, \&
  {Bestenlehner}}]{Sanyal_15}
{Sanyal}, D., {Grassitelli}, L., {Langer}, N., \& {Bestenlehner}, J.~M. 2015,
  \aap, 580, A20

\bibitem[{{Springmann}(1994)}]{Springmann_94}
{Springmann}, U. 1994, \aap, 289, 505

\bibitem[{{Springmann} \& {Puls}(1998)}]{Springmann_98}
{Springmann}, U. \& {Puls}, J. 1998, in Astronomical Society of the Pacific
  Conference Series, Vol. 131, Properties of Hot Luminous Stars, ed.
  I.~{Howarth}, 286

\bibitem[{{Sundqvist} {et~al.}(2019){Sundqvist}, {Bj{\"o}rklund}, {Puls}, \&
  {Najarro}}]{Sundqvist_19}
{Sundqvist}, J.~O., {Bj{\"o}rklund}, R., {Puls}, J., \& {Najarro}, F. 2019,
  \aap, 632, A126

\bibitem[{{Sundqvist} {et~al.}(2018){Sundqvist}, {Owocki}, \&
  {Puls}}]{Sundqvist_18}
{Sundqvist}, J.~O., {Owocki}, S.~P., \& {Puls}, J. 2018, \aap, 611, A17

\bibitem[{{Todt} {et~al.}(2010){Todt}, {Pe{\~n}a}, {Hamann}, \&
  {Gr{\"a}fener}}]{Todt_10}
{Todt}, H., {Pe{\~n}a}, M., {Hamann}, W.~R., \& {Gr{\"a}fener}, G. 2010, \aap,
  515, A83

\bibitem[{{van Leer}(1979)}]{Van_Leer_79}
{van Leer}, B. 1979, Journal of Computational Physics, 32, 101

\bibitem[{{Vink} {et~al.}(2001){Vink}, {de Koter}, \& {Lamers}}]{Vink_01}
{Vink}, J.~S., {de Koter}, A., \& {Lamers}, H.~J.~G.~L.~M. 2001, \aap, 369, 574

\bibitem[{{Wolf} \& {Rayet}(1867)}]{Wolf_1867}
{Wolf}, C.~J.~E. \& {Rayet}, G. 1867, Academie des Sciences Paris Comptes
  Rendus, 65, 292

\bibitem[{{Xia} {et~al.}(2018){Xia}, {Teunissen}, {El Mellah}, {Chan{\'e}}, \&
  {Keppens}}]{Xia_18}
{Xia}, C., {Teunissen}, J., {El Mellah}, I., {Chan{\'e}}, E., \& {Keppens}, R.
  2018, \apjs, 234, 30

\bibitem[{{Yoon} {et~al.}(2012){Yoon}, {Gr{\"a}fener}, {Vink}, {Kozyreva}, \&
  {Izzard}}]{Yoon_12}
{Yoon}, S.~C., {Gr{\"a}fener}, G., {Vink}, J.~S., {Kozyreva}, A., \& {Izzard},
  R.~G. 2012, \aap, 544, L11

\end{thebibliography}

\end{document}